\documentclass[10pt,journal,compsoc]{IEEEtran}

\usepackage[utf8]{inputenc}
\usepackage{numprint}
\usepackage{longtable}
\usepackage{color}
\usepackage{amsmath,amssymb,amsthm}
\usepackage{physics}
\usepackage{hyperref}
\usepackage{graphicx}
\usepackage{multirow}

\DeclareMathOperator*{\argmax}{arg\,max}

\newcommand{\val}{\textsf{gval}}
\newcommand{\dval}{\textsf{dval}}

\newcommand{\dist}{\textsf{dist}}
\newcommand{\cnot}{\textsc{cnot}}
\newcommand{\Had}{\textsc{h}}
\newcommand{\len}{\textsf{len}}
\newcommand{\sahs}{\textsc{sahs}}
\newcommand{\sabre}{\textsc{sabre}}
\newcommand{\swap}{\textsc{swap}}
\newcommand{\fidls}{\textsc{FiDLS}}

\newcommand{\tket}{{\textsf{t}|\textsf{ket}\rangle}}

\newcommand{\rred}[1]{{\color{black}{#1}}}

\newcommand{\blue}[1]{{\color{black}{#1}}}

\newtheorem{example}{Example}
\newtheorem{remark}{Remark}
\begin{document}
%
\title{Qubit Mapping Based on Subgraph Isomorphism and Filtered Depth-Limited Search}
%
%
%
%

\author{Sanjiang~Li,
        Xiangzhen~Zhou,
        Yuan~Feng
\IEEEcompsocitemizethanks{\IEEEcompsocthanksitem Sanjiang Li and Yuan Feng are with Centre for Quantum Software and Information (QSI), Faculty of Engineering and Information Technology, University of Technology Sydney, NSW 2007, Australia.\protect\\
E-mail: $\{$sanjiang.li, yuan.feng$\}$@uts.edu.au
\IEEEcompsocthanksitem Xiangzhen Zhou is with State Key Lab of Millimeter Waves, Southeast University, Nanjing 211189, China and Centre for Quantum Software and Information, University of Technology Sydney, NSW 2007, Australia.}
}

\markboth{Qubit mapping (Final): September~2020}%
{Shell \MakeLowercase{\textit{et al.}}: Bare Demo of IEEEtran.cls for Computer Society Journals}


\IEEEtitleabstractindextext{%
\begin{abstract}
    Mapping logical quantum circuits to Noisy Intermediate-Scale Quantum (NISQ) devices is a challenging problem which has attracted rapidly increasing interests from both quantum and classical computing communities. This paper proposes an efficient method by (i) selecting an initial mapping that takes into consideration the similarity between the architecture graph of the given NISQ device and a graph induced by the input logical circuit; and (ii) searching, in a {filtered and depth-limited way,} a most useful {\swap} combination that makes executable as many as possible two-qubit gates in the logical circuit. The proposed circuit transformation algorithm can significantly decrease the number of auxiliary two-qubit gates required to be added to the logical circuit, especially when it has a large number of two-qubit gates. For an extensive benchmark set of 131  circuits and IBM's current premium Q system, viz., IBM Q Tokyo, our algorithm needs, in average, 0.3801 extra two-qubit gates per input two-qubit gate, while the corresponding figures for three state-of-the-art algorithms are 0.4705, 0.8154, and 1.0066  respectively.
\end{abstract}

}

\maketitle

\IEEEdisplaynontitleabstractindextext
\IEEEpeerreviewmaketitle

\IEEEraisesectionheading{\section{Introduction}\label{sec:introduction}}
\IEEEPARstart{S}{ince} Shor's exciting quantum algorithms for solving integer factorisation and discrete logarithm \cite{Shor94}, many quantum algorithms have been proposed that could offer an exponential speed-up when compared with best classical algorithms. These include in particular the HHL algorithm for solving systems of linear equations \cite{HHL} and other machine learning algorithms derived from HHL (cf. \cite{Biamonte+17} for a  summary). Typically, the implementation of these algorithms requires quantum computers with millions of qubits which are perhaps still not available in the next two decades. On the other hand, IBM, Intel and Google have all announced their quantum devices with around 50-70 qubits recently. The Noisy Intermediate-Scale Quantum (NISQ) era seems coming in reality. Despite that quantum error correction is not yet available in the near future, 
\blue{quantum supremacy has recently been demonstrated in Google's 53-qubit quantum processor Sycamore \cite{GoogleQsupr}.}

\begin{figure}[t]
	\centering
	\includegraphics[width=0.3\textwidth]{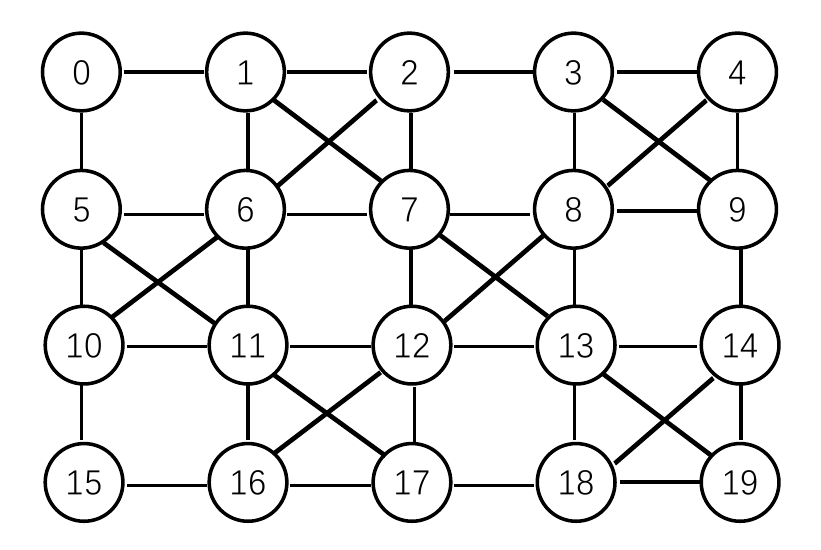}
	\caption{The architecture graphs for IBM Q Tokyo.}
	\label{fig:ibmq20}
\end{figure}

There is yet another gap between theoretical research on quantum algorithms and their implementation on realistic quantum devices. When designing quantum algorithms, typically, the quantum circuit model allows multi-qubit gates to act on any set of qubits without restriction. This is, however, not the case in realistic NISQ devices, which have ``limited number of qubits, limited connectivity between qubits,
restricted (hardware-specific) gate alphabets, and 
limited circuit depth due to noise" \cite{Khatri+19}. In the superconducting devices of IBM, Google, and Rigetti, only single and special two-qubit gates (like \cnot\ or \textsc{cz}) are supported. Even worse, these two-qubit gates can only be implemented between neighbouring qubits. For example, Fig.~\ref{fig:ibmq20} shows the architecture graph of IBM's current premium quantum system IBM Q Tokyo (also known as IBM Q20), which supports elementary single-qubit gates and two-qubit \cnot\ gates and a \cnot\ gate can be implemented only between qubits which are connected by an (undirected) edge. In order to use these NISQ devices, the desired quantum functionality in an ideal quantum circuit should be  \emph{transformed} or \emph{mapped} so that the underlying coupling constraints imposed by these quantum devices are satisfied.    

More precisely, in order to implement an ideal quantum circuit on an NISQ device like IBM Q Tokyo, we need to address two issues. The first is to  decompose the desired functionality (arbitrary quantum gates) into elementary operations that can be directly applied on the NISQ device. This issue has already been properly addressed in several works \cite{AmyMMR13,MatsumotoA08,WilleSOD13}. The second issue, known as \emph{qubit mapping} or \emph{circuit transformation}, is to map or route the qubits in the ideal quantum circuit to qubits of the quantum device so that the coupling constraints imposed by the quantum device are satisfied and thus the two-qubit gates in the ideal quantum circuit are executable. 

In the past several years, the qubit mapping problem has attracted rapidly increasing interests from both classical and quantum computing communities, see, e.g., \cite{MaslovFM08,SaeediWD11,Siraichi+18,Venturelli+18,Murali+19,ZulehnerPW18,LiDX19,ChildsSU19,CowtanDDKSS19,Zhou+19}
and references therein. Given an arbitrary quantum circuit and an NISQ device, the task of qubit mapping is to construct automatically a quantum circuit with the same functionality which can be immediately implemented in the NISQ device. For ease of presentation, we call the quantum circuit to be transformed a \emph{logical} circuit and call a circuit that is implementable in the NISQ device a \emph{physical} circuit. Similarly, we call qubits in a logical (physical) circuit logical (physical) qubits. We assume that gates in the input logical circuit are already decomposed into elementary gates supported by the NISQ device. In particular, each gate in the logical circuit involves at most two qubits. Naturally, we also assume that the number of logical qubits is not larger than the number of physical qubits.


It is not difficult to find such a solution for an input logical circuit. Indeed, we can start with an arbitrary initial mapping and then execute one two-qubit gate per round (note single-qubit gates can be executed directly) by inserting several \swap\ operations to transform the current mapping to a mapping that can execute the current two-qubit gate.  The challenge is if we can find a solution with minimal overhead in terms of the number of \swap\ gates inserted. This is crucial for the success of quantum computing as a large number of extra two-qubit gates will significantly accumulate the error of the output physical circuit. 

Finding an optimal solution for the qubit mapping problem is often very difficult. Indeed, it is NP-complete \cite{Siraichi+18} to decide if an input logical circuit can be transformed into an equivalent physical circuit {using up to $k$ {\swap} gates} for a fixed integer $k>0$. Several previous works use off-the-shelf tools like dynamic programming \cite{Siraichi+18}, SAT solvers \cite{SaeediWD11}, temporal planners \cite{Venturelli+18}, Integer Linear Programming (ILP) \cite{Almeida+19}, satisfiability modulo theory (SMT) solvers \cite{Murali+19}. In worst cases, all these approaches take time exponential in the number of qubits. 

Many other works devise specialised heuristic search algorithms for solving the qubit mapping problem. For example, Zulehner, Paler and Wille \cite{ZulehnerPW18} partitioned the input logical circuit into layers and introduced an $A^*$ search algorithm. When combined with a lookahead scheme and a dedicated method for selecting the initial mapping, their algorithm performs much better than IBM's own solution. However, this $A^*$-approach takes time exponential in the number of qubits. Li, Ding and Xie \cite{LiDX19} proposed a search algorithm based on reverse traversal, which is polynomial in the number of qubits and works very well for small circuits with less than a hundred two-qubit gates. Their algorithm, called \sabre,  outperforms the $A^*$-approach \cite{ZulehnerPW18} with exponential speedup and comparable or better results on various benchmarks. Childs, Schoute and Unsal \cite{ChildsSU19} also proposed efficient methods that attempt to minimise the circuit depth or size overhead and have worst-case time complexity polynomial in the sizes of the input circuit and the architecture graph. To this end, they decomposed the problem into two subproblems: qubit movement and qubit placement. The first subproblem considers how to transform the current mapping to a selected next mapping by imposing \swap\ gates on edges in the architecture graph, while the second subproblem gives method to compute the next mapping. 
In another work, Cowtan et al. \cite{CowtanDDKSS19} described a solution implemented in the platform-independent compiler $\tket$. They also partitioned the input circuit into layers and selected the  {\swap} which can maximally reduce the diameter of the subgraph composed of all pairs of qubits in the current layer. We address their algorithm as the Cambridge algorithm henceforth. In \cite{Zhou+19}, we designed a new qubit mapping algorithm, called \sahs\ in this paper, which uses simulated annealing for constructing an initial mapping and searches the next mapping by using a heuristic function that reflects the variable influence of gates in different layers. Empirical results show that \sahs\ outperforms both \sabre\ and the Cambridge algorithm by a large margin. {Initial mappings of \sahs\ are, however,} computed non-deterministically by simulated annealing, which is sometimes unstable and runs slowly when the circuit size is large \cite{Zhou+19}.      

In this paper, we propose a new search algorithm based on subgraph isomorphism and filtered depth-limited search. The idea is to construct a graph $G$ from the input circuit which is isomorphic to a subgraph of $\mathcal{AG}$, the architecture graph of the given NISQ device, and select any embedding from $G$ to $\mathcal{AG}$ as the initial mapping. Starting from this initial mapping, we then, step by step, construct the physical circuit while removing executable gates from the logical circuit. If the current mapping can execute some gates in the front layer of the logical circuit, we remove them from the logical circuit and properly append them to the current physical circuit; if there are no executable gates in the front layer, then we need to insert \swap\ gates and obtain a new mapping so that some two-qubit gates in the front layer can be executed. To select a good next mapping, we {tend to} exhaustively search all possible combinations of \swap\ operations such that the number of executable two-qubit gates per \swap\ is maximised. As selecting the best \swap\ combination is expensive, we fix $k>0$ and only consider combinations of at most $k$ {\swap}s. The search process could be further sped up if we `filter' those {\swap}s which do not interact with gates in the first several layers of the circuit. Such filters are designed and used in our algorithm. 

\blue{While less efficient when compared with the Cambridge algorithm, the search process of our algorithm is
polynomial in all relevant parameters} and can significantly reduce the number of {\swap}s required to transform the input logical circuit. Indeed, empirical evaluation shows that our algorithm can often reduce by  half the number of {\swap}s required when compared with \sabre, if the input logical circuit has hundreds or more two-qubit gates. Similar empirical evaluations also show that our algorithm is significantly better than the Cambridge algorithm \cite{CowtanDDKSS19} and our \sahs\ algorithm \cite{Zhou+19} in terms of the size of the output circuits.

The remainder of this paper is organised as follows. In Sec.~\ref{sec:background}, we recall some background of quantum computation and quantum circuits, and describe and analyse our algorithm in Sec.~\ref{sec:approach}. Detailed empirical evaluation on IBM Q Tokyo is reported in Sec.~\ref{sec:eval}. \blue{We further discuss in Sec.~\ref{sec:discussion} possible extensions, scalability, efficiency and effectiveness of our approach and report more evaluation results on three large architectures.} The last section concludes the paper with discussions on directions for future research.

\section{Backgrounds}
\label{sec:background}
In this section, after a brief introduction of quantum gates and quantum circuits, we describe the dependency graph associated to a logical circuit and show how to partition the logical circuit into layers by using the dependency graph. 

\subsection{Quantum Gates and Quantum Circuits}

\begin{figure*}[t]
	\centering
	\begin{tabular}{lr}
	\includegraphics[width=0.25\textwidth]{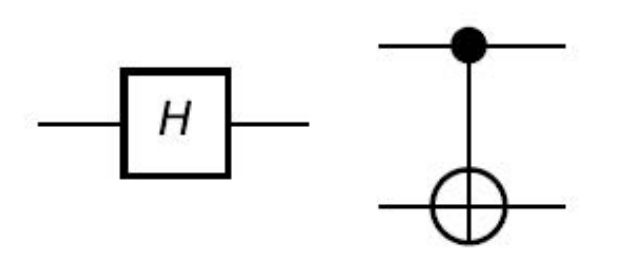}
	& 
	\includegraphics[width=0.35\textwidth]{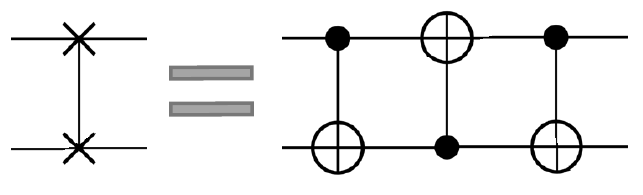}
	\end{tabular}
	\caption{Hadamard, {\cnot} and {\swap}  gate (from left to right).}
	\label{fig:gates}
\end{figure*}

Qubit is the counterpart of bit in quantum  computation. While a `classical' bit can only be in one of two states, viz., 0 and 1, a qubit can be in the superposition state $\ket{\psi}=\alpha \ket{0} + \beta\ket{1}$ of the two basis states, $\left| 0 \right\rangle$ and $\left| 1 \right\rangle$,
where $\alpha,\beta\in\mathbb{C}$ are probability amplitudes satisfying ${\left| \alpha  \right|^2} + {\left| \beta  \right|^2} = 1$. For example, $\ket{+}=\frac{1}{\sqrt{2}}(\ket{0}+\ket{1})$ and $\ket{-}=\frac{1}{\sqrt{2}}(\ket{0}-\ket{1})$ are two superposition states.
The success of quantum computation partially lies in ingenious use of \rred{quantum superposition}.

Quantum computation is realised by applying quantum gates on qubits. Complex, multi-qubit gates can be decomposed into elementary single or two-qubit gates. In fact, any quantum gate can be approximated to arbitrary accuracy using a fixed set of single-qubit gates and {\cnot} gates \cite{Barenco+95}.
Fig.~\ref{fig:gates} illustrates three very useful gates: Hadamard gate {\Had}, {\cnot} gate and {\swap}  gate. Hadamard gate is a single-qubit gate which can evenly mix the basis states to produce a superposed one. Precisely, {\Had} maps $\ket{0}$ to $\ket{+}$ and $\ket{1}$ to $\ket{-}$. {\cnot} and {\swap} are both two-qubit gates, i.e., they operate on two qubits. A {\cnot} gate flips the target qubit (indicated graphically with $\oplus$) if and only if the control qubit (indicated graphically with a black dot $\bullet$) is in state $\left| 1 \right\rangle$, while a {\swap}  gate exchanges the states of the two qubits operated. Precisely, {\cnot} maps $\ket{a}\ket{b}$ to $\ket{a}\ket{a\oplus b}$ and {\swap} maps $\ket{a}\ket{b}$ to $\ket{b}\ket{a}$ for $a,b\in \{0,1\}$. Most NISQ devices do not support \swap\ gates directly and, if this is the case, we may implement a \swap\ gate by three \cnot\ gates (see Fig.~\ref{fig:gates} (right)).

Quantum circuits are the most commonly used model to describe quantum algorithms, which consist of input qubits, quantum gates, measurements and classical registers \cite{Rodney14}. 
As only input qubits and quantum gates are relevant in the qubit mapping problem, in this paper, we represent a quantum circuit simply as a pair $\left( {Q,C} \right)$, where $Q$ is the set of involved qubits and $C$ a sequence of quantum gates.


\subsection{Dependency Graph and Front Layer}
\label{sec:front_layer}
\begin{figure*}
    \centering
    \begin{tabular}{lr}
	\includegraphics[width=0.5\textwidth]{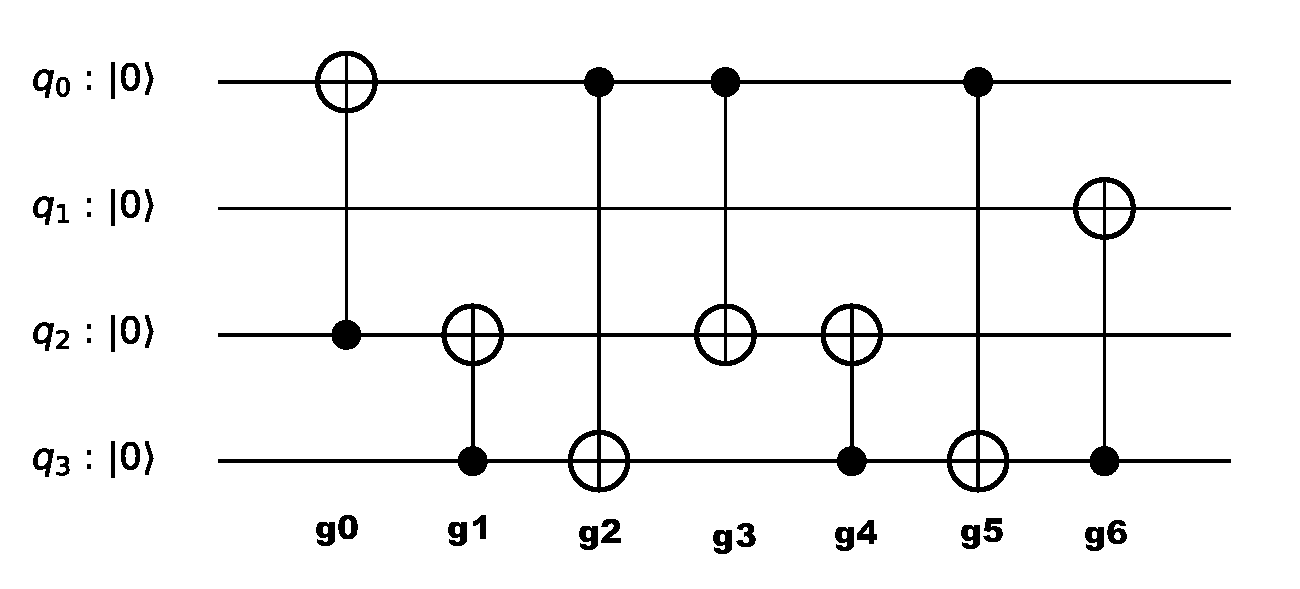}
	& 
    \includegraphics[width=0.13\textwidth]{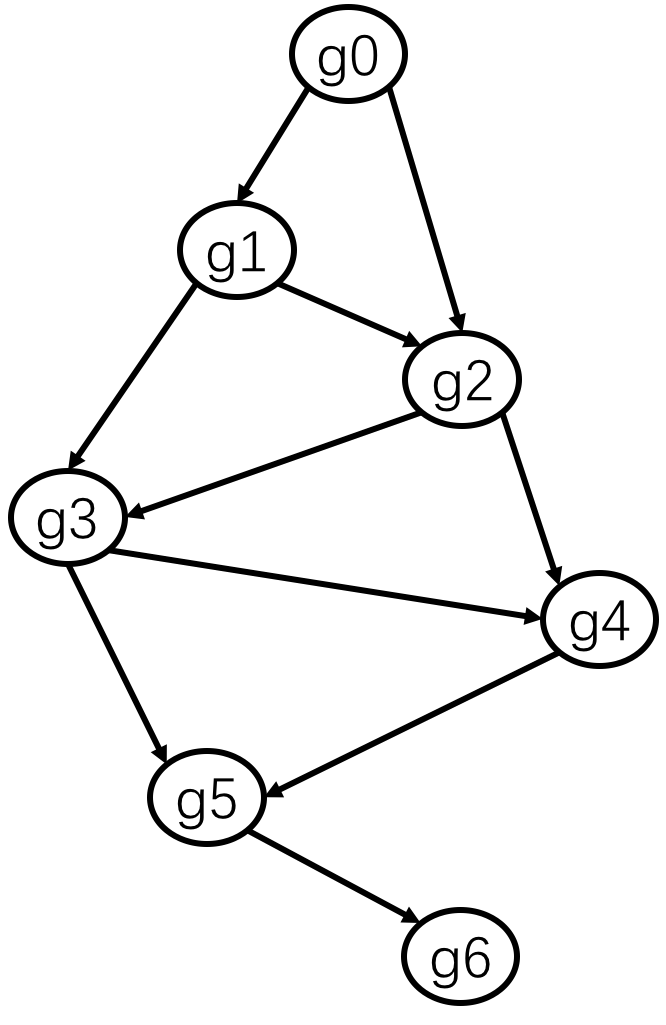}
    \end{tabular}
    \caption{\rred{A logical quantum circuit with only {\cnot} gates (left) and its dependency graph (right)}.}
    \label{fig:DG}
\end{figure*}

Two-qubit gates in a logical circuit $LC = (Q,C)$ are not independent in general. We say a two-qubit gate $g_1$ \emph{depends} on another two-qubit gate $g_2$ if the latter must be executed before the former. This happens when $g_2$ is in front of $g_1$ in $C$ and they share a common qubit, or when $g_1$ depends on a two-qubit gate $g_3$ which depends on $g_2$. \rred{For clarity, we say $g_1$ \emph{directly depends} on $g_2$ if  $g_2$ is in front of $g_1$ in $C$ and they share a common qubit and there are no other gates between them which share the same common qubit.}

For a logical circuit $LC = (Q,C)$, we construct a directed acyclic graph (DAG), called the \emph{dependency graph}, to characterise the \rred{direct} dependency between two-qubit gates in $LC$ \cite{LiDX19,ChildsSU19}. Each node of the dependency graph represents a two-qubit gate and each directed edge the \rred{direct} dependency relationship from one two-qubit gate to another. The front layer of $LC$, denoted $\mathcal{F}(LC)$ or $\mathcal{L}_0(LC)$, consists of all two-qubit gates
in $LC$ which have no parents in the dependency graph. The second layer $\mathcal{L}_1(LC)$ is then the front layer of the circuit obtained from $LC$ by deleting all gates in $\mathcal{F}(LC)$. Analogously, we can define the $k$-th layer  $\mathcal{L}_k(LC)$ of $LC$ for all $k\geq 0$.

\begin{example}\label{ex:cir}
Consider the logical circuit $LC = (Q,C)$ shown in Fig.~\ref{fig:DG} (left), where
\begin{align*}
    Q &= \{q_0,q_1,q_2,q_3\}, \\  
    C &= ( g_0 \equiv \langle q_2, q_0\rangle, g_1 \equiv \langle q_3,  q_2\rangle, g_2 \equiv \langle q_0, q_3\rangle, g_3 \equiv \langle q_0, q_2\rangle, \\
    & \quad\quad\quad g_4 \equiv \langle q_3, q_2\rangle, g_5 \equiv \langle q_0, q_3\rangle, g_6 \equiv \langle q_3, q_1\rangle ),
\end{align*} 
where $\langle q_2,q_0\rangle$, for example, denotes the {\cnot} gate in the logical circuit with $q_2$ being the control qubit and $q_0$ the target.

For this circuit, we have $\mathcal{F}(LC) = \{ g_0\}$, $\mathcal{L}_1(LC) = \{g_1\}$, $\mathcal{L}_2(LC) = \{g_2\}$, and $\mathcal{L}_3(LC) = \left\{ {{g_3}} \right\}$, and so on. From the dependency graph (showing in Fig.~\ref{fig:DG} (right)), we can see that, for example, gate $g_2$ can be executed only after $g_0$ and  $g_1$. 
\end{example}

\section{The Proposed Approach}
\label{sec:approach}
The main objective of qubit mapping is to transform an input logical circuit to a physical one with minimal size or depth so that the constraints imposed by the NISQ device are satisfied. To simplify the discussion, we only consider the connectivity constraints for two-qubit gates as specified by the architecture graph. This means that single-qubit gates have no effect in the circuit transformation process. Furthermore, we make the following assumptions:\footnote{The occurrences of {\cnot} gates may be replaced by \textsc{cz} gates when, e.g., a Rigetti device or Google's Sycamore is used.} 
\begin{enumerate}
    \item The NISQ device supports all single-qubit gates and \cnot\ gates; 
    \item The architecture graph of the NISQ device, $\mathcal{AG}$,  is an undirected graph;
    \item \cnot\ gates are the only two-qubit gates in the input logical circuit. 
\end{enumerate}

From now on and as in Example~\ref{ex:cir}, we write a \cnot\ gate simply as a pair $\langle q,q'\rangle$, where $q$ and $q'$ are the control and target qubits, respectively. We call the {\cnot} gate $\langle q',q\rangle$ the \emph{inverse} of $\langle q,q'\rangle$. \label{page:inverse_CNOT}

Let $\mathcal{AG}=(V,E)$ be the undirected architecture graph  of the NISQ device we are given, where $V$ is the set of physical qubits and $E$ the set of edges along which {\cnot} gates can be performed. Recall that an edge $e$ in an undirected graph is an unordered pair of  endnodes $p,q$ of $e$. In the following, we write $e$ simply by $\{p,q\}$, i.e., the set of its two endnodes.


\blue{Let $LC = (Q, C)$ be a logical circuit  with $|Q| \le |V|$. Suppose $C$ consists of only {\cnot} gates after removing all single-qubit gates. We need to construct a physical circuit $PC = (V, C^p)$ which contains only {\cnot} gates and satisfies:
\begin{itemize}
\item [1.] it is functionally equivalent to $LC$ after adding back all single-qubit gates accordingly, and 
\item [2.] it respects the connectivity constraints imposed by $\mathcal{AG}$, i.e., $\{q,q'\} \in E$ for any {\cnot} gate $\langle q,q'\rangle$ in $C^p$.
\end{itemize}
}

It is easy to find a physical circuit that satisfies the above conditions, but the real challenge is to find one with \emph{minimal} size or depth, which is NP-complete in general \cite{Siraichi+18}. 
In this paper, we modify the input logical circuit stepwise by inserting auxiliary  {\swap} operations (each implemented with three {\cnot} gates as in Fig.~\ref{fig:gates} (right)) until the logical circuit is transformed into a physical circuit that can be executed on the NISQ device. To evaluate the effectiveness of qubit mapping algorithms, we use the sizes of the output circuits, i.e., the total number of its two-qubit gates.

\subsection{Qubit Mapping}
\label{sec:qubit_mapping}

In each step, qubits in the logical circuit are mapped or allocated to physical qubits in the NISQ device. Mathematically, a (partial) qubit mapping is a (partial) function $\tau$ from $Q$ to $V$ such that $\tau(q)=\tau(q')$ if and only if $q=q'$ for any $q,q'\in Q$. We say a partial qubit mapping is \emph{complete} if it is defined for every $q$ in $Q$. \blue{A physical qubit $v$ is \emph{occupied} or \emph{allocated} if $\tau$ maps some logical qubit $q$ to $v$. Otherwise, we say $v$ is unoccupied.} The mapping may change in consecutive steps of the transformation which is determined by the inserted auxiliary \swap\ operations.

Given a logical circuit $LC$ and a mapping $\tau$, a {\cnot} gate $g=\langle q,q'\rangle$ in $LC$ is said to be \emph{satisfied} by $\tau$, or $\tau$ satisfies $g$, if $\{\tau(q),\tau(q')\}$ is an edge in $\mathcal{AG}$. Furthermore, $g$ is \emph{executable} by $\tau$ if it appears in the front layer of $LC$ and $\tau$ satisfies $g$.
If this is the case, we remove $g$ from $LC$ and append a {\cnot} gate $\tau(g) := \langle \tau(q),\tau(q')\rangle$ to the end of the physical circuit. This process is called the \emph{execution} of $g$.

For two physical qubits $v,v'$ in $\mathcal{AG}$, we write $\dist_{\mathcal{AG}}(v,v')$ for the distance (i.e., the length of a shortest path) from $v$ to $v'$ in $\mathcal{AG}$. 

For any mapping $\tau$ and any two-qubit gate $g=\langle q, q'\rangle$, the \emph{physical distance} between the two qubits $q,q'$ in $g$ under mapping $\tau$, \blue{written $\dist_{ph}(q,q',\tau)$, is defined as
\begin{itemize}
\item [-] the distance between $\tau(q)$ and $\tau(q')$ in $\mathcal{AG}$, i.e., $\dist_{\mathcal{AG}}(\tau(q),\tau(q'))$, if both $\tau(q)$ and $\tau(q')$ are defined,
\item [-]  
the shortest distance between $\tau(q)$ or $\tau(q')$ to all unoccupied physical qubits if only one of $\tau(q)$ and $\tau(q')$ is defined, and 
\item [-] the shortest distance between two unoccupied qubits if neither $\tau(q)$ nor $\tau(q')$ is defined.
\end{itemize}
}
Apparently, a gate $g=\langle q,q'\rangle$ in the front layer is executable by $\tau$ 
iff both $\tau(q)$ and $\tau(q')$ are defined and the physical distance between $q$ and $q'$ under $\tau$ is 1.

\begin{figure*}
    \centering
    \begin{tabular}{ccc}
	\includegraphics[width=0.16\textwidth]{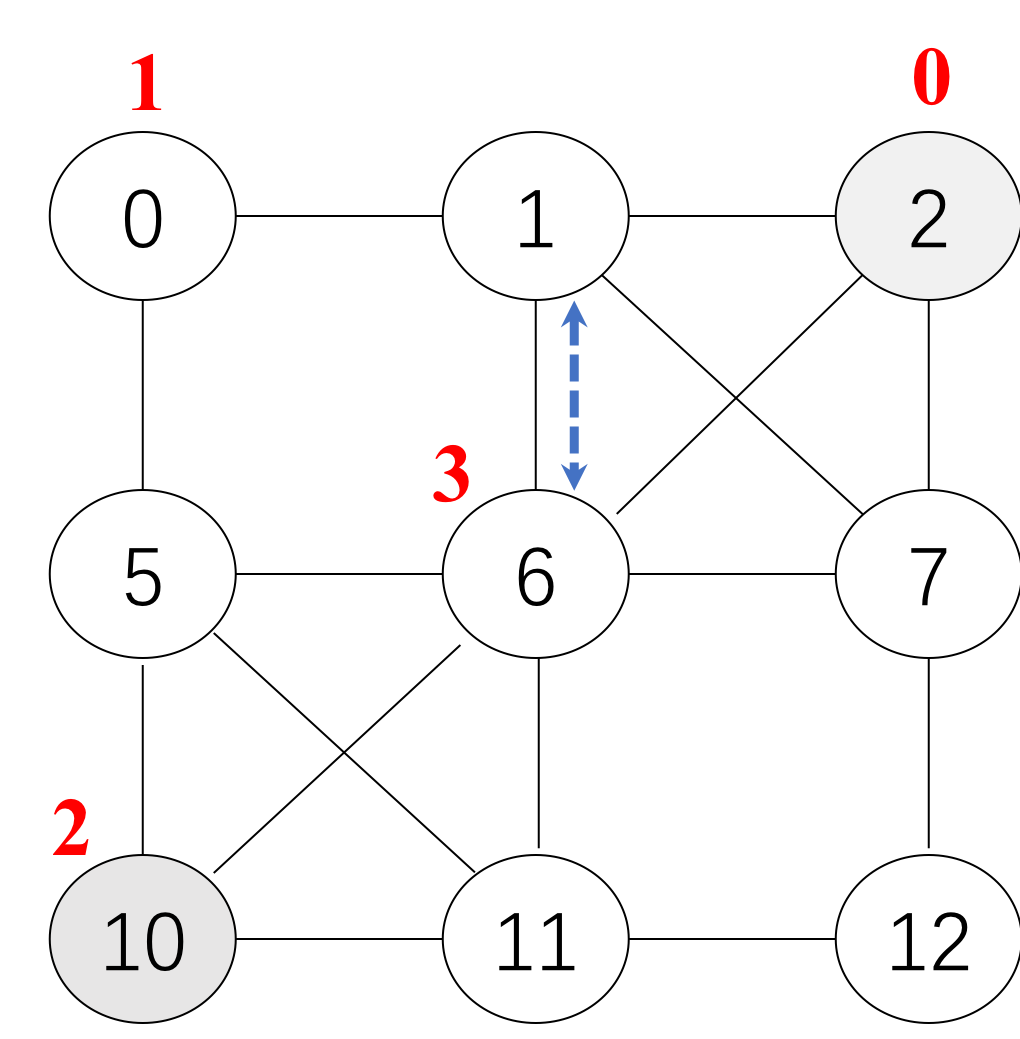}
	&
	\includegraphics[width=0.16\textwidth]{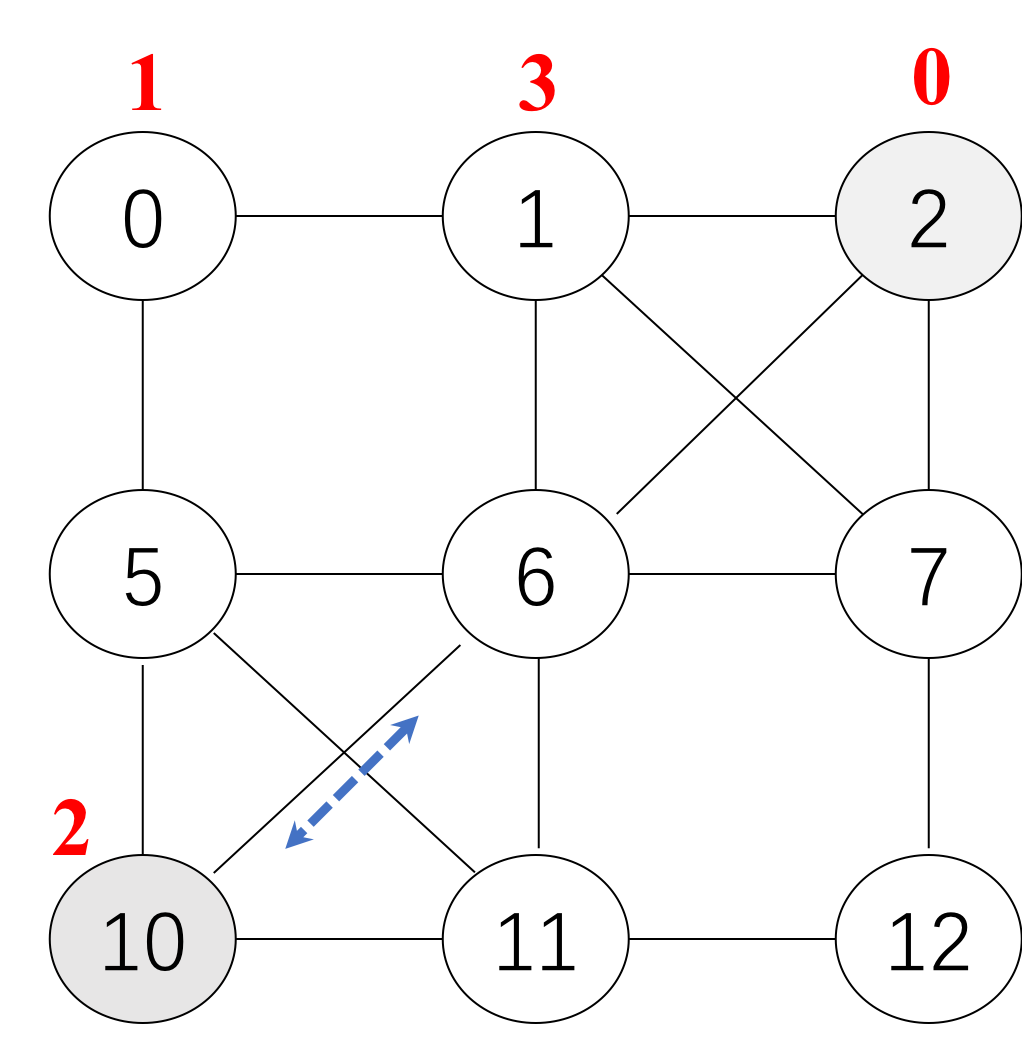}
    &
    \includegraphics[width=0.15\textwidth]{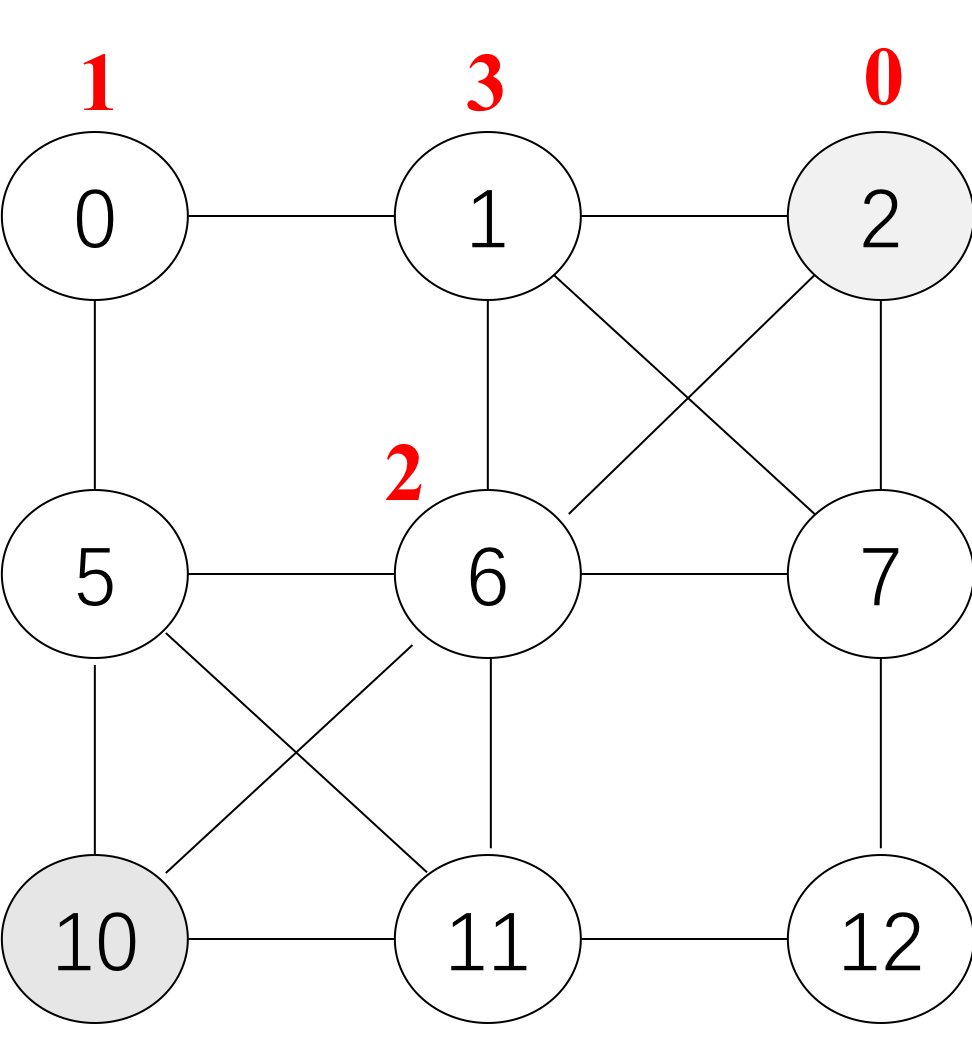}\\
    $\tau_1$  & $\tau_2$  & $\tau_3$  
    \end{tabular}
    \caption{Qubit mappings $\tau_{i}: \{q_0,q_1,q_2,q_3\} \to V$ for $i=1,2,3$, where $\tau_2$ is obtained from $\tau_1$ by \swap(1,6) and $\tau_3$ is obtained from $\tau_2$ by \swap(6,10), where $\swap(1,6)$ is a shorthand for $\swap(v_1,v_6)$.}
    \label{fig:map}
\end{figure*}
\begin{example}[Example~\ref{ex:cir} cont'd]
Consider the logical circuit $LC$ shown in Fig.~\ref{fig:DG}. Let $\tau_1:Q\to V$ be the mapping specified by $\tau_1(q_0)=v_2$, $\tau_1(q_1)=v_0$, $\tau_1(q_2)=v_{10}$, $\tau_1(q_3)=v_6$, see Figure~\ref{fig:map} (left). Then, because $\tau_1(q_2)=v_{10}$ and $\tau_1(q_0)=v_2$ and $\dist_{\mathcal{AG}}(v_2,v_{10})=2$, we can see that $g_0 \equiv \langle q_2, q_0\rangle$ is not executable by $\tau_1$ in IBM Q Tokyo. However, $g_0$ is executable by the mapping $\tau_3$ shown in Figure~\ref{fig:map} (right) and, indeed, every gate in $LC$ is satisfied by $\tau_3$. 
\end{example}

\subsection{Initial Mapping}
\label{sec:initial_mapping}

An initial mapping can be constructed step by step or selected arbitrarily or computed from a dedicated subroutine. Zulehner et al. \cite{ZulehnerPW18} tested both arbitrary initial mappings and initial mappings evolved from an empty mapping. Their experimental evaluation showed that, in general, the latter approach has better performance. In \cite{LiDX19}, Li et al. proposed to use an initial mapping that takes the whole input circuit into consideration. Starting from a randomly generated mapping $\tau_0$, they first take this mapping as the initial mapping and apply it on the input circuit $(Q,C)$, the obtained final mapping $\tau_{1}$ is then used as the initial mapping and applied to the inverse circuit\footnote{\rred{Note the ``inverse" here has a different meaning as the ``inverse" {\cnot} gate defined in page~\ref{page:inverse_CNOT}.}} $(Q,C^{inv})$, and lastly, the obtained final mapping $\tau_f$ is selected as the initial mapping for their main algorithm \sabre. Their approach was demonstrated as consistently better than the $A^*$ search algorithm in \cite{ZulehnerPW18}. In \cite{Zhou+19}, we proposed \sahs, which uses simulated annealing to search for the best initial mapping that fits well with the input logical circuit and empirical evaluation there shows that, when compared with the naive mapping that sends $q_i$ to $v_i$, \sahs\ works significantly better with the initial mapping obtained by simulated annealing.    



In this section we show how to obtain a good initial mapping by matching a particular graph induced by the input circuit with the architectural graph.



Suppose $LC=(Q,C)$ is the input logical circuit. We first construct an undirected graph $\mathcal{G}_{\textsf{circ}}(C)=(Q,E_{\textsf{circ}})$ on $Q$, where $\{q,q'\}$ is an edge in $E_{\textsf{circ}}$ if either $\langle q,q'\rangle$ or its inverse $\langle q',q\rangle$ is in $C$. If $\mathcal{G}_{\textsf{circ}}(C)$ happens to be isomorphic to a subgraph of the architecture graph $\mathcal{AG}$, then the qubit mapping problem is solved by constructing an \rred{(arbitrary)} isomorphic embedding $\tau$ from $\mathcal{G}_{\textsf{circ}}(C)$ to $\mathcal{AG}$. For NISQ devices, which have up to several hundreds qubits, this can be solved by, for example, the VF2 algorithm \cite{CordellaFSV04}. 
If $\mathcal{G}_{\textsf{circ}}(C)$ is not isomorphic to a subgraph of $\mathcal{AG}$, then we may select a maximal  sub-circuit $C_{\textsf{top}}$ of $C$ such that 
\begin{itemize}
    \item [(i)] $C_{\textsf{top}}$ is a front section of $C=\{g_i\in C \mid 1\leq i\leq n\}$ w.r.t. the dependency graph of $C$, i.e., a two-qubit gate $g_i$ is in $C_{\textsf{top}}$ only if all two-qubit gates $g_j$ on which $g_i$ depends are in $C_{\textsf{top}}$; 
    \item [(ii)] the graph $\mathcal{G}_{\textsf{circ}}(C_{\textsf{top}})$ is isomorphic to a subgraph of $\mathcal{AG}$; and 
    \item [(iii)] the graph $\mathcal{G}_{\textsf{circ}}(C_{\textsf{top}}\cup\{g_{i^*}\})$ is not isomorphic to a subgraph of $\mathcal{AG}$ for any $g_{i^*}$ that is in the front layer of $C\setminus C_{\textsf{top}}$. 
\end{itemize}
We call $\mathcal{G}_{\textsf{circ}}(C_{\textsf{top}})$ a \emph{top subgraph} (\emph{topgraph} for short) of $\mathcal{G}_{\textsf{circ}}(C)$. Let $\tau_{\textsf{top}}$ be an isomorphic embedding from $\mathcal{G}_{\textsf{circ}}(C_{\textsf{top}})$ to $\mathcal{AG}$. We select $\tau_{\textsf{top}}$ as the initial mapping, which \rred{satisfies} all gates in $C_{\textsf{top}}$. \rred{Note that $\tau_{\textsf{top}}$ might not be a complete mapping from $Q$ to $V$.}

\begin{example}[Example~\ref{ex:cir} cont'd]
For the circuit $LC$  in Example~\ref{ex:cir}, we have $Q=\{q_0,q_1,q_2,q_3\}$ and  $E_{\textsf{circ}} = \{\{q_0,q_2\},\{q_0,q_3\},\{q_2,q_3\}, \{q_1,q_3\}\}$. Clearly, $\mathcal{G}_{\textsf{circ}}(C)$ is isomorphic to a subgraph of $\mathcal{AG}$ and such an isomorphism is specified by the qubit mapping $\tau_3$ in Fig.~\ref{fig:map} (right).
\end{example}

Note that $\tau_{\textsf{top}}$ often does not take the whole circuit into consideration. We propose another method for constructing the initial mapping that considers the whole circuit. For a logical circuit $LC=(Q,C)$, \rred{we introduce a weight function $\omega$ which assigns a weight on each edge of $E_{\textsf{circ}}$ (the edge set of the undirected graph $\mathcal{G}_{\textsf{circ}}(C)$ defined above) such that $\omega(\{q,q'\})$ is the number of gates $g_i$ in $C$ with $g_i=\langle q,q'\rangle$ or $g_i=\langle q',q\rangle$. Let $E^w_{\textsf{circ}}=\{e_1,e_2,...,e_n\}$ where $\omega(e_1) \geq \ldots \geq \omega(e_n)$.}
%
We then construct a subgraph $\mathcal{G}^*=(Q,E^*)$ of $\mathcal{G}_{\textsf{circ}}(C)$ which is isomorphic to a subgraph of $\mathcal{AG}$ as follows. We start by letting $E^*=\{e_1\}$ and then consider the next edge $e_2$. In general, suppose we have decided if $e_i$ should be put into $E^*$ or not for all $i<k$ for some $k\leq n$ and the current subgraph $\mathcal{G}^*$ is isomorphic to some subgraph of $\mathcal{AG}$. We consider $e_{i+1}$. If putting $e_{i+1}$ into $E^*$ will make $\mathcal{G}^*$ non-isomorphic to any subgraph of $\mathcal{AG}$, we skip this edge; otherwise, we put $e_{i+1}$ into $E^*$ and update $\mathcal{G}^*$, which is still isomorphic to some subgraph of $\mathcal{AG}$. If $i+1<n$, we continue to consider $e_{i+2}$ till there is no edge left in $E^w_{\textsf{circ}}$. In this way, we obtain a subgraph $\mathcal{G}^*$ of $\mathcal{G}_{\textsf{circ}}(C)$ that is isomorphic to some subgraph of $\mathcal{AG}$. The sum of weights of edges in $\mathcal{G}^*$, though not necessary the largest, is sufficiently large among all subgraphs of $\mathcal{G}_{\textsf{circ}}(C)$ that are isomorphic to some subgraph of $\mathcal{AG}$. Using the VF2 algorithm, we can find an embedding $\tau_{\textsf{wgt}}$ which embeds $\mathcal{G}^*$ into $\mathcal{AG}$. \rred{Again, we note that $\tau_{\textsf{wgt}}$ might be a partial mapping from $Q$ to $V$.}

In the following, we call $\tau_{\textsf{top}}$ the \emph{topgraph initial mapping} and call $\tau_{\textsf{wgt}}$ the \emph{wgtgraph initial mapping} of $LC$. Besides these two initial mappings, we also introduce a method for evolving an initial mapping from the empty mapping. 
Similar idea was used by Zulehner et al. \cite{ZulehnerPW18}, while we extend a partial mapping only when necessary, i.e., when the thus extended mapping can execute a two-qubit gate in the current front layer or it can reduce the \emph{minimum} physical distance (cf. Sec.~\ref{sec:qubit_mapping}) between qubits in a two-qubit gate in the current front layer. This mapping extension technique is also used when  $\tau_{\textsf{top}}$ or $\tau_{\textsf{wgt}}$ is incomplete. 

\subsection{Fixed-Depth Heuristic Search}
\label{sec:search}

In most search-based algorithms for the qubit mapping problem, a heuristic function is used to select an  action (i.e., a {\swap} or a sequence of {\swap}s) which can maximally reduce the sum or the minimum of the distances between the two qubits in the {\cnot} gates of the front layer and, sometimes, the lookahead layer. 

For each edge $e=\{v,v'\}$ in $\mathcal{AG}$ there is an associated \swap\ operation, written $\swap(e)$, which swaps the states on $v$ and $v'$. More precisely, suppose $\tau$ is the current mapping and $\tau(q)=v$, $\tau(q')=v'$. Then $\swap(e)$ transforms $\tau$ into a new mapping $\tau'$ such that $\tau'(q)=v'$, $\tau'(q')=v$, and $\tau'(q^*)=\tau(q^*)$ for $q^*\not\in\{q,q'\}$. In case if $\tau(q)$ is not defined and $\tau(q')=v'$, then $\tau'(q')$ is not defined and $\tau'(q)=v'$. The case when $\tau(q')$ is not defined and $\tau(q)=v$ is analogous. If both are undefined, then $\tau'=\tau$. We often write ${\swap(e)}\circ\tau$ for $\tau'$.

\begin{example}[Example~\ref{ex:cir} cont'd]
For the three qubit mappings in Fig.~\ref{fig:map}, we have $\tau_2= \swap(1,6)\circ \tau_1$ and $\tau_3 = \swap(6,10)\circ \tau_2$.
\end{example}

In this section, we propose a new heuristic function which measures how \emph{efficient} the mapping can execute gates in the logical circuit. For convenience and by abuse of terminology, we say a {\cnot} gate not in the front layer is executable by a mapping $\tau$ if the gate itself and all {\cnot} gates it depends on are satisfiable by $\tau$. 


Starting with a selected initial mapping $\tau^0$, we write $s^0=(\tau^0,PC^0,LC^0)$ for the initial state of the search process, where $LC^0$ is obtained by removing all {\cnot} gates $\langle q,q'\rangle$ that are executable by $\tau^0$ from $LC$, and $PC^0$ is obtained by adding the corresponding {\cnot} gates $\langle \tau^0(q),\tau^0(q')\rangle$ in an empty physical circuit.  Step by step, we select an action $a$ from $\mathcal{S}$, the set of sequences of {\swap}s on $\mathcal{AG}$ and enforce all {\swap}s in $a$ one by one to get the next mapping (and the next state) till there are no gates left in the logical circuit.

Suppose $s^i=(\tau^i, PC^i, LC^i)$ is the current state and all gates that are executable by $\tau^i$ are already removed from $LC^i$. For a sequence $a=(\swap_1,\swap_2,...,\swap_\ell)$ of {\swap}s on $\mathcal{AG}$, we define a value function 
\begin{align}
    \label{eq:val_of_swaps}
    \val(\tau^i,a) & = \frac{\mbox{number of gates executable by $\tau'$}}{\len(a)\times 3},  
\end{align}
where $\tau'$ is the mapping obtained by enforcing {\swap}s in  action $a$ one by one on $\tau^i$ and $\len(a) = \ell$ is the number of {\swap}s in $a$. Recall each {\swap} is implemented by three {\cnot} gates (see Fig.~\ref{fig:gates} (right)). 

Our action set consists of all sequences of {\swap}s on $\mathcal{AG}$ and we select any one with \rred{the} maximal value, i.e., we select $a^*$ from 
\begin{align}
    \label{eq:next_action}
     \argmax_{a\in \mathcal{S}}\ \val(\tau^i,a).
\end{align}
 After selecting $a^*$, we enforce on $\tau^i$ {\swap}s in $a^*$ one by one and obtain the next mapping $\tau^{i+1}$. Then we remove all gates that are executable by $\tau^{i+1}$ from $LC^i$ and write $LC^{i+1}$ for the resulted logical circuit. In the meanwhile, we append to $PC^i$ three {\cnot} gates (as in Fig.~\ref{fig:gates} (right)) for each {\swap} in $a^*$, and a \cnot\ gate $\langle \tau^{i+1}(q),\tau^{i+1}(q')\rangle$ for each \cnot\ gate $\langle q, q'\rangle$ removed from $LC^{i}$. In this way, we obtain $PC^{i+1}$ and the next state $s^{i+1}=(\tau^{i+1},PC^{i+1},LC^{i+1})$.

Apparently, considering all sequences of {\swap}s is inefficient. In practice, we propose to consider actions with up to $k$ {\swap}s for some fixed $k\geq 1$. In particular, for IBM Q Tokyo, we select $k=3$, which reflects a good compromise between efficiency and effectiveness.

\begin{example}[Example~\ref{ex:cir} cont'd] \label{ex:cir-5}
Suppose $\tau_1$ is the qubit mapping which maps $q_0, q_1,q_2,q_3$ to, respectively, $v_2, \rred{v_0}, v_{10}, v_6$, see Fig.~\ref{fig:map} (left). As the front layer contains only  $g_0=\langle q_2,q_0\rangle$, which is not executable by $\tau_1$, there are no gates in $LC$ that can be executed by $\tau_1$. Examining all sequences of up to 3 {\swap}s, the four best actions are as follows:
\begin{itemize}
    \item $a_1= (\swap(1, 6), \swap(6, 10))$, which can execute all 7 gates in $LC$;
    \item $a_2= (\swap(5, 6), \swap(2, 6))$, which can execute all 7 gates in $LC$;
    \item $a_3= (\swap(6, 7), \swap(6, 10))$, which can execute all but the last gate in $LC$;
    \item $a_4= (\swap(6, 11), \swap(2, 6))$, which can execute all but the last gate in $LC$.
\end{itemize}
The fifth best action contains 3 {\swap}s. Thus $a_1$ and $a_2$ are optimal actions, with the optimal value $\val(\tau_1,a_1)=\val(\tau_1,a_2)=7/6$.

\end{example}

\subsubsection{Heuristics used in related works}

Now it is a good time to compare our heuristic function with those used in the related works. 

Zulehner et al. \cite{ZulehnerPW18} selected the action that results in a mapping which can execute all gates in the front layer and the lookahead layer. The action consists of a sequence of {\swap}s and is selected by using $A^*$ search and the following heuristics:
\begin{align*}
    h(\tau^i) & = \sum_{\langle q,q'\rangle \in \mathcal{L}_0\cup\mathcal{L}_1} \bigg\{ 3\times \big(\dist_{\mathcal{AG}}(\tau^i(q),\tau^i(q'))-1 \big) \bigg\}, 
\end{align*}
where $\mathcal{L}_0$ and $\mathcal{L}_1$ are the first two layers of the current logical circuit and, for any two-qubit gate $\langle q,q'\rangle$, $\dist_{\mathcal{AG}}(\tau^i(q),\tau^i(q'))$
is the distance from  $\tau^i(q)$ to $\tau^i(q')$ in $\mathcal{AG}$. The heuristic cost is not admissible and thus an optimal action is not guaranteed. Moreover, the worst-case time complexity of this $A^*$ search algorithm is exponential in the number of logical qubits.

Childs et al. \cite{ChildsSU19} selected the action which can maximally reduce the total distance between qubits in the \cnot\ gates in the current front layer, i.e., 
\begin{align}
    \label{eq:h-Childs}
    R(\tau^i) & = \sum_{\langle q,q'\rangle \in \mathcal{L}_0} \dist_{\mathcal{AG}}(\tau^i(q),\tau^i(q')). 
\end{align}
Their algorithm is polynomial in all relevant parameters but its performance is not directly compared with the $A^*$ algorithm in \cite{ZulehnerPW18}.

To overcome the inefficiency of the $A^*$ search algorithm, several researchers (see, e.g., \cite{LiDX19,CowtanDDKSS19}) proposed to select a \emph{single} {\swap} each time. Their methods are more efficient than the $A^*$-approach when processing logical circuits with more than 15 qubits. In \cite{LiDX19}, Li et al. designed a heuristic cost function that can reduce the sum of distances between the two qubits in each two-qubit gate in the front (and the lookahead) layers. Analogously, Cowtan et al.  \cite{CowtanDDKSS19} used a heuristic cost function that can reduce the diameter of the subgraph composed of all qubits in the two-qubit gates of the front layer. In the evaluation section, we will see that the efficiency of these  algorithms is partially achieved at the cost of the quality of the output physical circuit. 

In the \sahs\ algorithm \cite{Zhou+19}, we introduced a heuristic function that supports weight parameters to reflect the variable influence of gates in different layers. In each step, instead of selecting the action with the minimal cost, \sahs\ selects the {\swap} which has the best consecutive {\swap} to apply.

\subsection{Optimisation and Fallback}
\label{sec:fallback&opt}

Considering all sequences of up to $k$ {\swap}s is still not efficient for devices with a medium to large architecture graph. Let $\tau$ be the current mapping and $\mathcal{L}_i$ the current $i$-th  $(0\leq i \leq k)$ layer. Write $Q_i$ for the set of logical qubits in $\mathcal{L}_i$. It's natural not to consider {\swap}s that do not interact with gates in the first one or several layers. This idea was used for edges in the front layer in, e.g., \cite{ZulehnerPW18,LiDX19,Zhou+19}.

For an edge $e=\{v,v'\}$, if neither $\tau^{-1}(v)$ nor $\tau^{-1}(v')$ is in $Q_0$, then swapping $v$ and $v'$ does not reduce the minimum distance between qubits in a two-qubit gate in the current front layer, viz. $\mathcal{L}_0$. Therefore, it is reasonable to introduce the following filter for selecting an action $a=(e_1,e_2,...,e_\ell)$ with at most $k$ {\swap}s:


\begin{enumerate}
    \item [1.] $Q_0$-filter: We say $a=(e_1,e_2,...,e_\ell)$ is a $Q_0$-{plausible} action if, for any edge $e_j=\{v,v'\}$ of $a$, we have either $\tau_{j-1}^{-1}(v)$ or $\tau_{j-1}^{-1}(v')$ is in $Q_0$, where $\tau_0 \equiv \tau$ and $\tau_j$ is obtained from $\tau_{j-1}$ by enforcing $\swap(e_j)$ for $1\leq j\leq \ell$.
\end{enumerate}
Similarly, we could look ahead and introduce $Q_i$-filter for $0<i<k$.
\begin{enumerate}
    \item [2.] $Q_i$-filter: 
    We say $a=(e_1,e_2,...,e_\ell)$ is a $Q_i$-{plausible} action if, for any edge $e_j=\{v,v'\}$ of $a$ with $j>i$, we have either $\tau_{j-1}^{-1}(v)$ or $\tau_{j-1}^{-1}(v')$ is in $Q_i$, where $\tau_0 \equiv \tau$ and $\tau_j$ is obtained from $\tau_{j-1}$ by enforcing $\swap(e_j)$ for $1 \leq j\leq \ell$.
\end{enumerate}
The above $Q_i$-filter could be weakened by requiring that either $\tau_{i-1}^{-1}(v)$ or $\tau_{i-1}^{-1}(v')$ is in (a subset of) $Q_0\cup Q_1 \cup \cdots \cup Q_{i}$.  In our evaluation, we  used various combinations of $Q_0$ and $Q_1$ filters and the results are very promising (see Sec.~\ref{sec:eval}).

It should be stressed that, sometimes, $Q_0$-filter may `filter' out optimal actions. 
\begin{example}[Example~\ref{ex:cir} cont'd]
Note that $Q_0=\{0,2\}$ and $\tau_1(q_0)=v_2$, $\tau_1(q_2)=v_{10}$. Each $a_i$ for $1\leq i\leq 4$ in Example~\ref{ex:cir-5} is not $Q_0$-plausible. Thus our algorithm with $Q_0$-filter cannot find an optimal action. In fact, the following $Q_0$-plausible action is selected by our algorithm  
\begin{itemize}
    \item $a_5= (\swap(2, 7), \swap(1,6), \swap(6, 10))$.
\end{itemize}
This action (see Fig.~\ref{fig:map}) can execute all 7 gates in $LC$ and has $\val(\tau_1,a_5)=7/9$.
\end{example}

\blue{Another type of filters is also introduced in our algorithm. Let $\tau$ be the current mapping and $\mathcal{L}_i$  the $i$-th layer of the current logical circuit. Recall the notion of physical distance defined in Sec.~\ref{sec:fallback&opt} and assume that $\gamma \in [0,1]$ is a discount factor and $s \geq 0$. For two mappings $\tau_1$ and $\tau_2$, we say $\tau_1$ is $s$-better than $\tau_2$ if 
$\widehat{R}_s(\tau_1) < \widehat{R}_s(\tau_2)$, where for $i=1,2$ we have
\begin{align} \label{eq:Rs-better}
 \widehat{R}_s(\tau_i) & = \sum_{\ell = 0}^s \gamma^{\ell}\times \bigg(  \sum_{\langle q,q'\rangle \in \mathcal{L}_\ell} \dist_{ph}(q,q',\tau_i) \bigg). 
\end{align}
Intuitively, $\tau_1$ being $s$-better than $\tau_2$ implies that it is easier to transform all gates in the first $s$ levels if we start from $\tau_1$ instead of $\tau_2$, or, in other words, the `distance' from $\tau_1$ to the first $s$ levels is shorter than that from $\tau_2$.}

\blue{Analogously, we define the $D_s$-filter as follows. 
\begin{enumerate}
    \item [3.] $D_s$-filter: 
    We say $a=(e_1,e_2,...,e_\ell)$ is a $D_s$-{plausible} action if $\tau_{j-1}$ is \emph{not} $s$-better than $\tau_j$ for any $1\leq j \leq \ell$, where $\tau_0 \equiv \tau$ and $\tau_j$ is obtained from $\tau_{j-1}$ by enforcing $\swap(e_j)$ for $1 \leq j\leq \ell$.
\end{enumerate}
Note that if $\gamma=0$ then $D_s$-filter is the same as $D_0$-filter.
}

\subsubsection*{Fallback}
For a prefixed positive integer $k$, it is possible that, in some cases, no sequence of {\swap}s with length $\leq k$ can lead to a mapping which can execute any \cnot\ gate in the current front layer. If this is the case, we use the following natural fallback:

\noindent {\bf Fallback:} 
Select any {\swap} that can reduce $FB(\tau)$, the minimum distance between qubits in a two-qubit gate in the current front layer, which is formally defined as follows:
\begin{align}
    \label{eq:fallback}
    FB(\tau) & = \min_{\langle q,q'\rangle \in \mathcal{L}_0} \dist_{\mathcal{AG}}(\tau(q),\tau(q')). 
\end{align}

It is worth noting that, for our experiments on IBM Q Tokyo and an extensive set of logical circuits, the fallback is {rarely} activated.

\subsection{Complexity Analysis}
\label{sec:complexity}
\blue{From above we can see that our QCT algorithm has two independent processes: 
\begin{itemize}
\item [1.] the initial mapping construction process, and 
\item [2.] the search process. 
\end{itemize}
The construction of the topgraph and wgtgraph initial mappings requires determining if a graph is embeddable in another graph, which in the worst-case has time complexity exponential in the number of qubits.
Although it is an NP-complete problem, there are practical and efficient algorithms, say the VF2 algorithm \cite{CordellaFSV04}, 
which can quickly solve the subgraph isomorphism problem for graphs with several thousands nodes. Fortunately, the QCT problem in the NISQ era considers only graphs with up to 1000 nodes and thus, by using VF2, we can construct the topgraph and wgtgraph initial mappings in a reasonable time. 
Experiments on various architecture graphs with up to 361 nodes and circuits with up to 50 qubits and 15,000 \cnot\ gates confirm that the time consumption is acceptable. Please see Sec.~\ref{sec:scale} for  detailed empirical results and discussion.}


In the following we give a rough estimation of the complexity of the search process of our algorithm.

Suppose $LC=(Q,C)$ is a logical circuit and $\mathcal{AG}=(V, E)$  the architecture graph of a NISQ device. Write $|Q|$, $|V|$, and $|E|$ for, respectively, the cardinalities of $Q$, $V$, and $E$. Let $m$ be the number of \cnot\ gates in $C$, $diam$ and $\deg$ the diameter and maximum degree of $\mathcal{AG}$, respectively. 

We have the following simple observations:
\begin{itemize}
    \item The dependency graph of $LC$ can be computed in time linear in $m$, the number of \cnot\ gates in $C$.
    \item For any mapping $\tau$ and any logical circuit $LC$, we can identify (and remove from $LC$ as well as from its dependency graph) in time linear in $m$ the set of gates in $LC$ executable by $\tau$.
\end{itemize}

We first consider the ideal case when fallback is never activated during the search process. As described in Sec.~\ref{sec:search}, starting with a selected initial mapping $\tau^0$, step by step, we select an action $a$ consisting of up to $k$ {\swap}s on $\mathcal{AG}$ and enforce all {\swap}s in $a$ one by one to get the next mapping till there are no gates left in the logical circuit. Suppose $s^i=(\tau^i, PC^i, LC^i)$ is the current search state.  As there are at most $O(|E|^k)$ actions with up to $k$ {\swap}s, we can generate at most $O(|E|^k)$ different mappings from $\tau^i$. To select from these mappings the one which can execute the most gates in $LC^i$, we need time $O(|E|^k\cdot m)$ (cf. the second observation above). Because each step removes at least one {\cnot} from $LC$, in at most  $O(|E|^k\cdot m^2)$ time, we can execute all gates in $LC$.  \blue{This is often an overestimated upper bound. If any $Q$-filter is used, there are at most $|Q|^k\cdot \deg^k$ actions with up to $k$ {\swap}s, where $\deg$ is the maximum degree of $\mathcal{AG}$---which is 6 for IBM Q Tokyo and 4 for  grid-like AGs.}

Now, suppose fallback is activated. Since each activation of the fallback reduces by (at least) one the minimum distance between qubits in a {\cnot} gate in the current front layer (i.e., $FB(\tau)$ in Eq.~\ref{eq:fallback}), the whole search process activates the fallback procedure at most $m\times diam$ times. Note that each activation (see Eq.~\ref{eq:fallback}) needs to compute the shortest distance between the control and target qubits in a {\cnot} gate in the front layer of the current logical circuit and there are at most $|Q|/2$ {\cnot} gates in the front layer. Using Dijkstra's algorithm with lists, $FB(\tau)$ can be computed in time $O(|Q|\cdot |V|^2)$.\footnote{\blue{In practice, we precompute the distance between every two nodes of $\mathcal{AG}$ and store it in a table.}} Thus the total fallback on-cost is at most $O(|Q|\cdot |V|^2 \cdot m\cdot diam)$. 

Therefore, the overall time complexity of the search process is $O(|E|^k\cdot m^2+|Q|\cdot |V|^2 \cdot m\cdot diam)$. As $|Q| \leq |V| \leq |E|+1$ and $diam$ is usually very small when compared with $m$, the overall time complexity is bounded by $O(|E|^k\cdot m^2)$ if $k\geq 3$. In practice, this could be significantly reduced if we use $Q$-filters as the base in 
$|E|^k$ can be significantly reduced. 

As for the space complexity, in each state $s$, we maintain, besides the logical and physical circuits, the dependency graph of the current logical circuit and the set of plausible actions with up to $k$ {\swap}s. Thus the space complexity of the algorithm is bounded by $O(|E|^k+m)$. 

\section{Evaluation}
\label{sec:eval}

In this section, we compare our approach with the \sabre\ algorithm of Li et al. \cite{LiDX19}, \rred{the Cambridge algorithm of \cite{CowtanDDKSS19}}, and our qubit transformation algorithm \sahs\ based on simulated annealing and heuristic search \cite{Zhou+19}, which are three state-of-the-art algorithms for the qubit mapping problem on IBM Q Tokyo (see Fig.~\ref{fig:ibmq20}). Although we focus on a particular NISQ device in the evaluation, our approach is applicable to any undirected architecture graph, including Rigetti 16Q Aspen-4, IBM Q Rochester and Google's Sycamore\footnote{Note that Aspen-4 and Sycamore support \textsc{CZ} instead of \cnot.}. 
We use Python as our programming language and IBM Qiskit \cite{qiskit} as auxiliary environment.\footnote{Code available at
https://github.com/ebony72/FiDLS
} 
All experiments are conducted in a MacBook Pro with 3.1 GHz Intel Core i5 processor and 8GB memory. 

As for benchmark circuits, we consider all publicly available circuits evaluated in \cite{LiDX19} or \cite{CowtanDDKSS19}. Note that  only {\cnot} gates are considered in our comparison. For each individual circuit, we extract all its {\cnot} gates and use the thus reduced circuit as the input of our qubit mapping algorithm. We then compare the involved algorithms in terms of the number of auxiliary \cnot\ gates required. One may also use the following relative measure $R_{\cnot}$, first introduced in \cite{CowtanDDKSS19}, to compare different algorithms on a particular circuit:
\begin{align*}
    R_{\cnot} &= \frac{\mbox{ \#\cnot\ in the output physical circuit}} {\mbox{\#\cnot\ in the input logical circuit}}. 
\end{align*}
In order to compare algorithms evaluated over different benchmark sets of circuits, for any benchmark set $\mathcal{B}$ of quantum circuits, we define the following ${\cnot}$ index, written $I^\mathcal{B}_{\cnot}$, of an algorithm relative to $\mathcal{B}$ as the fraction of the total number of \cnot\ gates in all output physical circuits  over the total number of \cnot\ gates in all input logical circuits from $\mathcal{B}$, i.e.,
\begin{align*}
    I^{\mathcal{B}}_{\cnot} &= \frac{\sum_{LC \in \mathcal{B}}\mbox{\#\cnot\ in the output physical circuit of $LC$}} {\sum_{LC \in \mathcal{B}}\mbox{\#\cnot\ in $LC$}}. 
\end{align*}
For convenience, we write $\mathcal{B}_s$ and $\mathcal{B}_c$ for the benchmark sets of circuits used in \cite{LiDX19} and \cite{CowtanDDKSS19} respectively. Note that $\mathcal{B}_c$ contains {131} circuits that includes all the {23} circuits in  $\mathcal{B}_s$. For more precise comparison, we decompose $\mathcal{B}_c$ into three categories according to the number of {\cnot}s these benchmark circuits contain: small (0-99), medium (100-999) and large ($\geq 1000$). 

For all experiments reported in this paper, we fix the search depth $k$ as 3 and, if not otherwise specified, use $Q_0$-filter for the first {\swap} and $Q_1$-filter for all the other {\swap}s for filtering actions $a=(e_1,...,e_\ell)$ with at most 3 {\swap}s. \blue{We denote by Q01 this combination of $Q_0$ and $Q_1$ filters and use it as our default $Q$-filter. In addition, we also adopt the $D_0$-filter to exclude more less significant actions.}


\subsection{Comparison Among Different Initial Mappings} 

We first compare the two subgraph isomorphism related initial mappings (viz., the topgraph initial mapping $\tau_{\textsf{top}}$ and the wgtgraph initial mapping $\tau_{\textsf{wgt}}$) introduced in Section~3 with the empty mapping and the naive initial mapping (which maps $q_i$ to $v_i$ for each $q_i$ in $Q$). Table~\ref{tab:summary}  
summarises the results for all (small, medium, large) circuits in $\mathcal{B}_c$.
For each of these circuits and each initial mapping, the transformation can be completed within \blue{about 500 seconds} by using our algorithm. 

From Table~\ref{tab:summary}, we can see that the two isomorphism subgraph related initial mappings, $\tau_{\textsf{top}}$ and $\tau_{\textsf{wgt}}$ are significantly better than the empty initial mapping and the naive initial mapping for small and medium circuits, but the difference is not significant when large circuits are evaluated. This is not a surprise as the search heuristics plays a dominant role if the circuit has a large size. Note that if a logical circuit can be transformed into a physical circuit with zero overhead, our algorithm, when using either the topgraph or wgtgraph initial mapping, will very likely 
detect this.

Since the topgraph initial mapping is slightly better than the other three initial mappings, in the following, when compared with other algorithms, we always use the topgraph initial mapping.

\begin{table*}
    \centering
    \begin{tabular}{c|cccccccc|c}
   benchmarks    & \#circ. & topgr.i.m. & wghtgr.i.m. & empty i.m. & naive i.m. & \textsc{sahs} & Cambridge & \sabre\ & topgr.i.m.Q01x \\     \hline
    small  & 63 & 1.2521 &	1.3175	& 1.5041 &	1.8760	& 1.2619	&	1.5103  & 1.4547 & {\bf 1.2348}\\
    medium & 39 & 1.2886	& 1.3034	& 1.4074	& 1.5727 & 	1.3101	&	1.6854 & 1.6972 & {\bf 1.2460} \\
    large & 29 & 1.4280	& 1.4366 &	1.4504	 & 1.4422 & 	\blue{1.4763}	&	1.8211 & 2.0189 & {\bf 1.3847}	 \\
    all & 131 & 1.4231 &	1.4324	& 1.4497 &	1.4486	& \blue{1.4705}	&	1.8154 & 2.0066 & {\bf 1.3801}


    \end{tabular}
    \caption{Summary of the $I^{\mathcal{B}}_{\cnot}$-index of our algorithm with four different initial mapping constructing methods and \textsc{sahs} and Cambridge} \label{tab:summary}
\end{table*}





\subsection{Comparison with SABRE, Cambridge, and SAHS}

We then compare our algorithm with \sabre\  \cite{LiDX19} on the small benchmark set $\mathcal{B}_s$ of circuits used in \cite{LiDX19}. We use the topgraph initial mapping $\tau_{\textsf{top}}$. \blue{Recall that \sabre\ starts with a random initial mapping. For each circuit, we execute \sabre\ on our computer 5 times and take as the overhead the smallest number of added gates out of the 5 attempts and take its time consumption as the total time of the 5 attempts. The results are reported in Table~\ref{tab:Q20}, where we can see that the running time of \sabre\ is comparable with ours.}

\blue{Let $n_{sabre}$ and $n_{ours}$ be the numbers of \cnot\ gates added, respectively, by \sabre\ and by ours. The `Comparison' column of Table~\ref{tab:Q20} shows the ratio
$n_{ours}/n_{sabre}$, which is set as 1 if $n_{ours} = n_{sabre} = 0$. Apparently, the smaller the ratio is, the better our algorithm performs. From Table~\ref{tab:Q20} we can see that \blue{only one} circuit has ratio larger than 1.\footnote{Shortly we will analyse why our algorithm works bad on this circuit in Example~\ref{ex:counter-example}, Sec.~\ref{sec:effectiveness}.}}  For all circuits with more than \blue{500 {\cnot} gates, the ratio is at most 62\%}. In terms of the {\cnot} index, we have successfully decreased the index $I^{\mathcal{B}_s}_{\cnot}$ from $1+47808/50534=1.9461$ to $1+20790/50534=1.4114$.

We further compare our algorithm with the Cambridge algorithm of  \cite{CowtanDDKSS19} and our \sahs\ algorithm \cite{Zhou+19} on the large benchmark set $\mathcal{B}_c$, which contains  {131} circuits. A summary of the results in terms of the {\cnot} index is presented in Table~\ref{tab:summary}. 
It is worth stressing that the results of the Cambridge algorithm as presented in Table~\ref{tab:summary} are obtained without using postmapping optimisations. Precisely, we have removed the following codes from their algorithm:
\begin{itemize}
    \item `Transform.OptimisePhaseGadgets().apply(tkcirc)'
\item `Transform.OptimisePostRouting().apply(outcirc)'.
\end{itemize}
\blue{This is because the Cambridge algorithm was implemented in C++ and compiled, and the above postmapping optimisation codes are not directly transplantable to our algorithm. To provide a fair comparison,  we did not do any postmapping optimisations in our algorithm either. In Table~\ref{tab:summary}, we didn't include the time information of these algorithms. However, we note that Cambridge is super fast; it takes only 5.2 seconds to transform the 131 circuits in $\mathcal{B}_c$!}



\begin{table*}
    \centering
    \begin{tabular}{cccc|ccccc}
    \hline
    \begin{tabular}[c]{@{}c@{}} Circuit\\ Name\end{tabular} & 
    \begin{tabular}[c]{@{}c@{}} qubit \\ no. \end{tabular} &
    \begin{tabular}[c]{@{}c@{}}input \\ gate\end{tabular} &
    \begin{tabular}[c]{@{}c@{}} input \\ {\sc cnot} \end{tabular} &
    \begin{tabular}[c]{@{}c@{}} \sabre \\ added  \end{tabular}  &
    \begin{tabular}[c]{@{}c@{}} \sabre \\ time (s)  \end{tabular}  &
    \begin{tabular}[c]{@{}c@{}} topgraph \\ added    \end{tabular} &
    \begin{tabular}[c]{@{}c@{}} topgraph \\ time (s)    \end{tabular} &
    \small{Comp.} \\
    \hline
    4mod5-v1\_22      & 5     & 21    & 11      & 0     & 0.02		 & 0     & 0     & 1      \\
    mod5mils\_65      & 5     & 35    & 16      & 9     & 0.03		& 0     & 0      & 0      \\
    alu-v0\_27        & 5     & 36    & 17      & 3     & 0.03		& 9     & 0.07   & 3    \\
    decod24-v2 43     & 4     & 52    & 22      & 12    & 0.04		 & 0     & 0     & 0      \\
    4gt13\_92         & 5     & 66    & 30      & 21     & 0.05		& 0     & 0     & 0      \\
    ising\_model\_10  & 10    & 480   & 90      & 0    & 0.09	 & 0     & 0    & 1      \\
    ising\_model\_13  & 13    & 633   & 120     & 0     & 0.12		& 0     & 0     & 1      \\
    ising\_model\_16  & 16    & 786   & 150     & 0     & 0.23		& 0     & 0    & 1      \\
    qft\_10           & 10    & 200   & 90      & 48    & 0.17		& 39    & 1.6     & 0.81     \\
    qft\_16           & 16    & 512   & 240     & 171   & 0.57		& 153   & 21.06     & 0.89     \\
    rd84\_142         & 15    & 343   & 154     & 141   & 0.36		& 72   & 5.01     & 0.51      \\
    adr4\_197         & 13    & 3439  & 1498    & 1185  & 5.69		& 630   & 19.85     & 0.53     \\
    radd\_250         & 13    & 3213  & 1405   & 1092   & 4.98		& 555   & 28.21     & 0.51     \\
    z4\_268           & 11    & 3073  & 1343    & 1072  & 3.92		& 630   & 16.19    & 0.59     \\
    sym6\_145        & 7      & 3888  & 1701    & 1290  & 6.63		& 513 & 2.93      & 0.40     \\
    misex1\_241      & 15     & 4813  & 2100    & 1275  & 9.22	& 786 & 24.62     & 0.62     \\
    rd73\_252        & 10     & 5321  & 2319    & 2250  & 9.94		& 1095 & 9.72     & 0.49     \\
    cycle10\_2\_110  & 12     & 6050  & 2648    & 2406  & 14.32		& 1194 & 10.93    & 0.50     \\
    square\_root\_7  & 15     & 7630  & 3089    & 2403  & 19.06		& 1338 & 228.29     & 0.56     \\
    sqn\_258         & 10     & 10223 & 4459    & 4404  & 37.01		& 1578 & 17.73   & 0.36     \\
    rd84\_253        & 12     & 13658 & 5960    & 6291  & 39.57		& 2352  & 54.35   & 0.37     \\
    co14\_215        & 15   & 17936   & 7840    & 8946  & 34.77		& 4257  & 128.33   & 0.48     \\
    sym9\_193        & 11   & 34881   & 15232   & 14790 	& 406.19		& 5589  & 70.29     & 0.38     \\ \hline
    sum             & -     & 117289  & 50534  & 47808 & 593.01 & 20790 & 639.18 & \textbf{0.43}
    \end{tabular}
    \caption{Comparison of our algorithm with \sabre\ in  \cite{LiDX19} on IBM Q Tokyo. \blue{The numbers in the last column indicate the ratio of our added \cnot\ gates (with the topgraph initial mapping) against that of \sabre. 
    } 
    }
    \label{tab:Q20}
\end{table*}

From Table~\ref{tab:summary} we can see that, for the benchmark set $\mathcal{B}_c$, the $I^{\mathcal{B}_c}_{\cnot}$ index of our algorithm is $1.4231$, while the indices for \sahs\ and Cambridge are, respectively, 1.4705 and 1.8154. This shows that our algorithm can in average generate significantly better results than Cambridge and  \sahs. This is particularly true for large circuits which contain 1000 or more {\cnot} gates. \blue{When compared with \sahs, our algorithm is slightly better for small circuits, 2 points (1.2886 vs. 1.3101) better for medium circuits, and 4.8 points better for large circuits (1.4280 vs. 1.4763). Both algorithms are significantly better than Cambridge in all three categories.} 

In the above experiments, we used $Q_0$-filter for the first {\swap} and $Q_1$-filter (see Section~\ref{sec:fallback&opt}) for all the other {\swap}s when filtering actions with up to 3 {\swap}s. \blue{
If we use $Q_0$-filter for all {\swap}s, then the index becomes 1.4774, which is about 5 points inferior to the index 1.4231  reported in Table~\ref{tab:summary}. However, if we weaken $Q_1$-filter by using the qubits in the front layer and the lookahead layer, then the index could be further improved to \textbf{1.3801} from 1.4231. We denote by Q01x  \label{page:Q01x} this weakened combination of $Q_0$ and $Q_1$-filters. This better performance is achieved at the cost of relatively slower search process:  the total time consumption for the whole benchmark set is now 13,224 seconds, or 3.7 hours, while for six very large circuits (e.g., `mlp4\_245' with 16 qubits and 8232 {\cnot} gates), the transformation process requires \blue{900-1600} seconds.}

\section{\blue{Further Discussion}}
\label{sec:discuss}
\label{sec:discussion}
From the above evaluation, we can see that our algorithm has significant better performance on IBM Q Tokyo than state-of-the-art algorithms. In this section, we give a more detailed discussion on the effectiveness, extensibility and time efficiency of our algorithm. In particular, we report more experiments on three larger architecture graphs, viz., Sycamore (53 qubits), Rochester (53 qubits), and an artificial 19x19 Grid-like device with 361 qubits, called Q19x19.
\subsection{Extension}
Our filtered depth-limited approach actually can adopt  heuristic value functions other than Eq.~\ref{eq:val_of_swaps}. Indeed, we have implemented another value function based on the function specified in Eq.~\ref{eq:Rs-better}. The new heuristic value function is obtained by replacing ``number of {\bf g}ates executable by $\tau'$" in Eq.~\ref{eq:val_of_swaps} with $\widehat{R}_s(\tau^i)-\widehat{R}_s(\tau')$, the difference of the `{\bf d}istance' to the first $s$ levels of the current logical circuit from $\tau^i$ and $\tau'$. In this way we have a new value function
\begin{align}
    \label{eq:val_Rs_of_swaps}
    \dval_s(\tau^i,a) & = \frac{\widehat{R}_s(\tau^i)-\widehat{R}_s(\tau')}{\len(a)\times 3},  
\end{align}
where $\tau'$ is the mapping obtained by enforcing {\swap}s in  action $a$ one by one on $\tau^i$ and $\len(a) = \ell \leq k$ is the number of {\swap}s in $a$.  
Using this value function, we then select any action with the maximal $D$-value as our next action. To distinguish between the two implementations of our algorithm, we call the one using $\val$ as \fidls-G$s$, and the other using $\dval$ as \fidls-D$s$, where $s$ indicates the level of gates we considered. It is easy to see that \fidls-D$s$ has the same computational complexity as \fidls-G$s$.  


\subsection{Effectiveness} \label{sec:effectiveness}
While it works much better in average and especially on large circuits, our algorithm performs not better on several small or medium circuits. For example, the circuit `alu-v0$\_$27’ contains 5 qubits and 17 \cnot\ gates. To execute it on IBM Q Tokyo, our QCT algorithm needs to insert 3 (2, resp.) swaps if the topograph (wgtgraph, resp.) initial mapping is used, while both Cambridge and \sabre\ only require 1 swap.

This bad performance is perhaps due to three reasons. First, during the search process, our algorithm works in a greedy way and always tries to find the action with the best value. This, however, often leads to a series of optimal local transformations, which gives no guarantee on the optimality of the global transformation. Second, there are many different mappings which can embed a given graph into $\mathcal{AG}$. Our algorithm selects an arbitrary one. Selecting a better embedding by, for instance, comparing their $\widehat{R}_s$ value (cf.  Eq.~\ref{eq:Rs-better})
 may further improve the performance of our algorithm. Third, our selection of the initial mapping is greedy (cf. Sec.~\ref{sec:initial_mapping}). For example, when constructing the topgraph initial mapping, we select a maximal sub-circuit $C_{top}$ of $C$ whose corresponding graph  
$\mathcal{G}_{\textsf{circ}}(C_{\textsf{top}})$ is isomorphic to a subgraph of $\mathcal{AG}$. The following example shows that, however, this is not always a good choice.

\begin{figure}[t]
	\centering
    \begin{tabular}{ccc}
	\includegraphics[width=0.18\textwidth]{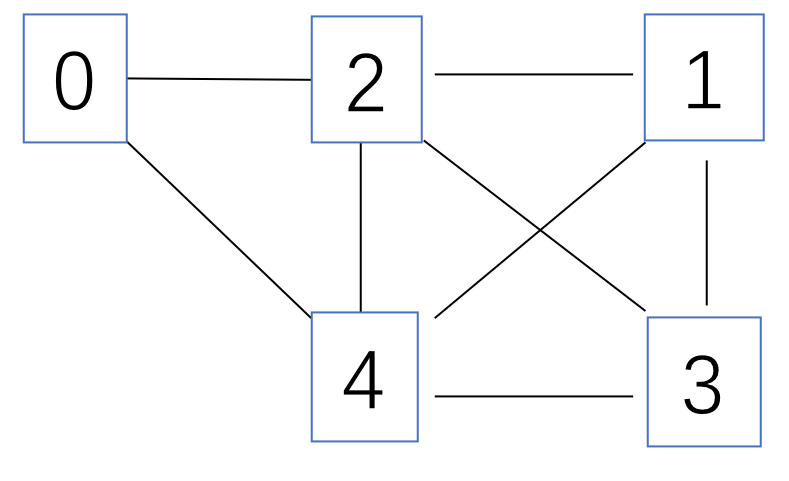}
	& &
	\includegraphics[width=0.2\textwidth]{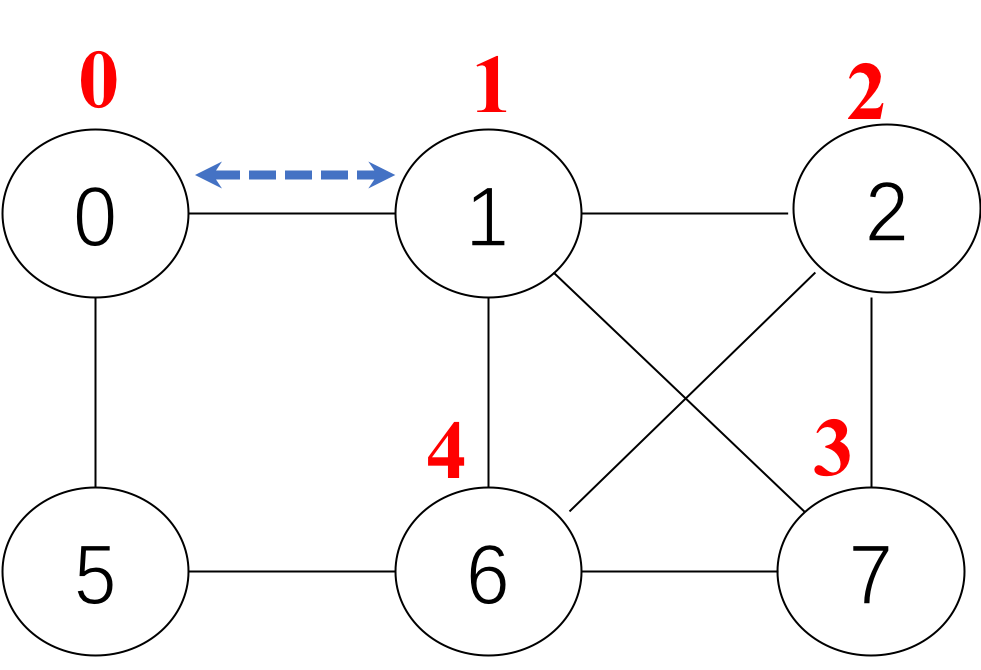}
	\end{tabular}
	\caption{The circuit graph of `alu-v0$\_$27’ (left) and a mapping (right).}
	\label{fig:cg65}
\end{figure}

\begin{example}\label{ex:counter-example}
Consider the circuit `alu-v0$\_$27’, which contains 5 qubits and the 17 \cnot\ gates in
\begin{align*}
  C=&\big(\langle 3, 4 \rangle , \langle 2, 1 \rangle, \langle 1, 3 \rangle, \langle 2, 1 \rangle, \langle 3, 2 \rangle, \langle 3, 1 \rangle, \langle 2, 1 \rangle, \langle 3, 2 \rangle, \langle 1,\\ & 3 \rangle, 
  \langle 2, 0 \rangle, \langle 0, 4 \rangle, \langle 2, 0 \rangle, \langle 4, 2 \rangle, \langle 4, 0 \rangle, \langle 2, 0 \rangle, \langle 4, 2 \rangle, \langle 0, 4 \rangle\big).  
\end{align*}
The circuit graph (Fig.~\ref{fig:cg65} (left)) is not embeddable in $\mathcal{AG}$, the architecture graph of IBM Q Tokyo (cf. Fig.~\ref{fig:ibmq20}). A part of $\mathcal{AG}$ is shown in the right of Fig.~\ref{fig:cg65}. Let $C_{top}$ be the sub-circuit of the first 10 gates in $C$. Then the circuit graph of $C_{top}$ is the topgraph of $C$ and can be embedded in $\mathcal{AG}$.  Let $\tau'$ be such an embedding. After removing gates in $C_{top}$, the rest gates (involving only qubits 0, 2, and 4) cannot be solved with one swap from $\tau'$. However, letting $\tau$ be the mapping as shown on the right of Fig.~\ref{fig:cg65}, we can see that $\tau$ solves only the first 9 gates but, after swapping 0 and 1, all the rest 7 gates can be solved.  
\end{example}



\subsection{Scalability and More on Time Complexity} \label{sec:scale}
In Sec.~\ref{sec:complexity} we have seen that the search process of our algorithm is polynomial in all relevant parameters. As we go deeper in the search tree, our algorithm becomes considerably slower than Cambridge. The significant decrease of the \cnot\ index shows that this is, however, worthwhile. 

In contrast with the search process, the initial mapping construction process relies on the efficiency of subgraph isomorphism algorithms, which have time complexity  exponential in the number of qubits in the circuit and the AG. We argue that this is not a serious problem for the following reasons. First, for the QCT problem in the NISQ era, only graphs with up to 1000 nodes are involved, which are manageable by existing algorithms like VF2 \cite{CordellaFSV04}. Second, the architecture graphs usually have very simple and regular (e.g., grid-like) topologies, which can be exploited to design customised efficient subgraph isomorphism algorithms. Last but not least, good approximate solutions can often do the job well. This is partially evidenced by the results summarised in Table~\ref{tab:summary}, where it shows that, for large circuits, the transformation results with an empty mapping are only slightly inferior to that using the selected topgraph initial mappings.   


To further evaluate the effectiveness and efficiency of our approach, we have experimented on IBM Q Rochester (53 qubits), Google’s Sycamore (53 qubits), and the artificial 19$\times$19 grid architecture graph Q19x19, which has 361 nodes and each node has at most 4 neighbours.  The results are summarised in Table~\ref{tab:4AGs-B131}, where we adopted the two value functions specified in Eqs.~\ref{eq:val_of_swaps} and ~\ref{eq:val_Rs_of_swaps}, respectively.  The two search methods are noted as \fidls-G and \fidls-D, respectively, where the latter uses the $D_2$-filter and the discount factor $\gamma$ in Eq.~\ref{eq:val_Rs_of_swaps} is fixed as 0.8 and all other parameters are same as in Sec.~\ref{sec:eval}. 

\begin{table}[thb]
    \centering
    \begin{tabular}{c|cccc}
   $\mathcal{B}_c$   & Tokyo & Sycamore & Rochester & Q19x19 \\     \hline
    topgr. constr. time  &73.9 &	26.8	& 26	& 38.9 \\
constr. time (max)    & 15.4	& 4.9 &	5.1	& 5.2	\\
    search time  (G) & 3820.8	&5594.0	&2166.7	&10533.8  \\
    search time (D) & 2388.2	&2228.6	&1535.1	&6669.4\\
         Cambridge time 
    &  5.2	&21.0	&19.2	&4231.3 \\ \hline
    I-index  (G) & {\bf 1.4231} &  2.7074	& 3.4835 & 2.7621\\	  
    I-index (D) & 1.8928 & 2.5929 	 & {\bf 3.1203} &  2.6408 \\  
    Cambridge I-index & 1.8154 & {\bf 2.5206}  & 3.136  &   {\bf 2.5095}	
    \end{tabular}
    \caption{Performance of our algorithm on four AGs and benchmark $\mathcal{B}_c$, where Cambridge is  implemented in C++ and times are in seconds.} \label{tab:4AGs-B131}
\end{table}

From Table~\ref{tab:4AGs-B131}, we can see that Cambridge is super fast. This is partially because it is implemented in C++, which is usually 10-100 times faster than the same Python program. 
Nevertheless, when the number of nodes of the AG goes from 20 to 361, the time consumption of our algorithm does not deteriorate too much while that of Cambridge increases more than 800 times. 

As circuits in the benchmark set $\mathcal{B}_c$ have only up to 16 qubits, we also tested another benchmark of circuits with large number of qubits. The benchmark\footnote{Available online at \url{https://github.com/CQCL/pytket/tree/master/examples/benchmarking/ChemistrySet}.},  written $\mathcal{B}_{bigQ}$ and first used in \cite{Cowtan_2020}, 
consists of a selection of circuits for Quantum Computational Chemistry generated using the Qiskit chemistry package \cite{qiskit}. $\mathcal{B}_{bigQ}$ contains 19 circuits with 20-50 qubits and up to 15000 \cnot\ gates. Results are summarised in Table~\ref{tab:4AGs-BbigQ}, where we can see that the running time of our algorithm on Q19x19 is comparable with that of Cambridge (implemented in C++). 

As for the effectiveness on the three large NISQ devices, our algorithm does not perform consistently better than Cambridge. For benchmark $\mathcal{B}_c$, which contains 131 circuits with up to 16 qubits, Table~\ref{tab:4AGs-B131} shows that \fidls-D and Cambridge perform better than \fidls-G. However, for benchmark $\mathcal{B}_{bigQ}$, Table~\ref{tab:4AGs-BbigQ} shows that our algorithm performs better than Cambridge on all three large NISQ devices, which is particular true for Sycamore and Q19x19. While we are not certain why this happens, it may be related to the following facts: (a) the AG of IBM Q Tokyo has larger average degree and smaller diameter than the other three large AGs, and (b) when searching for the best \swap\ action, Cambridge tends to keeps the occupied qubits staying together. The former fact seems especially favourable for \fidls-G while the latter fact seems very helpful for Cambridge when compiling circuits with small number of qubits on large AGs but this is not the case when compiling circuits with large number of qubits.



\begin{table}[thb]
    \centering    
    \begin{tabular}{c|cccc}
   $\mathcal{B}_{bigQ}$     & Sycamore & Rochester & Q19x19\\     \hline
    topgr. constr. time    & 2131.4	& 131.2	& 1254 \\
    constr. time (max) & 237.3	& 34.7	& 305.2\\
    search time (G)  & 556.8	& 713.6  & 714.9\\
    search time (D)  & 211.1	& 270.8  & 265.2\\
    Cambridge time & 2.5 & 3.0 & 516.6 \\ \hline
    I-index (G)  & {\bf 1.232} 	 &2.0977 &	1.3351 	\\
    I-index  (D)   & 1.3914	 &	{\bf 1.7131} & {\bf 1.3242} \\
    Cambridge I-index &1.4164  &  1.8443 &  	1.8439
    \end{tabular}
    
    \caption{Performance of our algorithm on large AGs and benchmark $\mathcal{B}_{bigQ}$, where Cambridge is  implemented in C++ and times are in seconds.} \label{tab:4AGs-BbigQ}

\end{table}
\begin{remark}
From Table~\ref{tab:4AGs-BbigQ}, we can see that the I-index results on Q19x19 are not always better than that on Sycamore. 
This implies that both Cambridge and our algorithm did not exploit the fact that the AG of Sycamore is embeddable into that of Q19x19 and, thus, every circuit that is executable on Sycamore can be executed on Q19x19 with a good initial mapping. This observation should be included in the implementation of every QCT algorithm. 
\end{remark}


\section{Conclusion}
\label{sec:conclusion}
We have proposed a new algorithm for qubit mapping based on subgraph isomorphism and filtered depth-limited search. Our algorithm, called \fidls, can significantly reduce the extra two-qubit gates required in the output circuit. If the input circuit can be executed directly, \fidls\ can very likely detect this. It seems that this nice property is not enjoyed by many other approaches. 

From our experimental results, we can see that, when the circuit has less than 1000 two-qubit gates, our subgraph isomorphism induced initial mappings are much better than empty mappings and naive mappings that assign the $i$-th qubit in the logical circuit to the $i$-th qubit in the quantum device. 
\blue{Experiment results on a large set of 131 benchmark circuits show that \fidls\ performs significantly better than state-of-the-art algorithms on IBM Q Tokyo, the architectural graph of which has a relatively large average node degree and smaller diameter. More experiments on three large NISQ devices with up to 361 qubits and the 131 benchmark circuits as well as 19 additional benchmark circuits with 20-50 qubits further demonstrated the scalability, efficiency, and effectiveness of our algorithm.}

\blue{Siraichi et al.  \cite{Siraichi+19} suggested to allocate qubit by combining subgraph isomorphism with token swapping. The idea is to partition the input circuit into a series of segments each of which can be executed by finding an appropriate mapping and then `glue' these mappings by token swapping. Experiment results in \cite{Siraichi+19} show that their algorithm can produce output circuits with 16\% less gates than \sabre\ on IBM Q Tokyo, which is not better than our results as reported in Table~\ref{tab:Q20}. This is perhaps due to that a series of optimal local mappings do not always lead to an optimal global one. See Example~\ref{ex:counter-example} for such a counter-example.}

A weighted graph like ours (see Section~\ref{sec:initial_mapping}) is also introduced in a recent work \cite{Lin+19}, where Lin, Anschuetz, and Harrow exploited spectral graph theory to qubit mapping. The performance of their algorithm is in general not better than the $A^*$ approach of \cite{ZulehnerPW18}. They also suggested to use their ``spectral mapper to provide an initial mapping" while its effectiveness needs further investigation.

It seems that we are still quite far from devising algorithms that could output circuits with nearly minimal overheads. Future work will investigate along the following directions: 
\begin{itemize}
    \item \blue{A better, possibly customised, subgraph isomorphism algorithm will improve the quality of our results. The approximate subgraph isomorphism algorithm proposed in \cite{Siraichi+19} seems a promising one.}
    
    \item Minimising depth or latency and circuit error is also important for qubit mapping. Although the number of CNOT gates in our output circuit is already smaller than the depth of the output circuit (only CNOT gates are counted) of some compared algorithm \cite{CowtanDDKSS19} (see \cite{Zhou+19} for a more detailed analysis), it will be nice if we can adapt our approach to address other or multiple optimisation objectives.
    
    \item Machine learning and deep learning algorithms may be designed to quickly select the best action in Eq.~\ref{eq:next_action}, especially when the search depth becomes large.
\end{itemize}

\bibliographystyle{IEEEtran}
\bibliography{qct_ref}

\begin{thebibliography}{10}
\providecommand{\url}[1]{#1}
\csname url@samestyle\endcsname
\providecommand{\newblock}{\relax}
\providecommand{\bibinfo}[2]{#2}
\providecommand{\BIBentrySTDinterwordspacing}{\spaceskip=0pt\relax}
\providecommand{\BIBentryALTinterwordstretchfactor}{4}
\providecommand{\BIBentryALTinterwordspacing}{\spaceskip=\fontdimen2\font plus
\BIBentryALTinterwordstretchfactor\fontdimen3\font minus
  \fontdimen4\font\relax}
\providecommand{\BIBforeignlanguage}[2]{{%
\expandafter\ifx\csname l@#1\endcsname\relax
\typeout{** WARNING: IEEEtran.bst: No hyphenation pattern has been}%
\typeout{** loaded for the language `#1'. Using the pattern for}%
\typeout{** the default language instead.}%
\else
\language=\csname l@#1\endcsname
\fi
#2}}
\providecommand{\BIBdecl}{\relax}
\BIBdecl

\bibitem{Shor94}
P.~W. Shor, ``Polynominal time algorithms for discrete logarithms and factoring
  on a quantum computer,'' in \emph{Proceedings of the First International
  Symposium on Algorithmic Number Theory}, 1994, p. 289.

\bibitem{HHL}
A.~W. Harrow, A.~Hassidim, and S.~Lloyd, ``{Quantum algorithm for linear
  systems of equations},'' \emph{Physical Review Letters}, vol. 103, no.~15, p.
  150502, 2009.

\bibitem{Biamonte+17}
J.~Biamonte, P.~Wittek, N.~Pancotti, P.~Rebentrost, N.~Wiebe, and S.~Lloyd,
  ``Quantum machine learning,'' \emph{Nature}, vol. 549, pp. 195 EP --, 09
  2017.

\bibitem{GoogleQsupr}
F.~Arute, K.~Arya, R.~Babbush, and et~al., ``Quantum supremacy using a
  programmable superconducting processor,'' \emph{Nature}, vol. 574, p.
  505–510, 2019.

\bibitem{Khatri+19}
S.~Khatri, R.~LaRose, A.~Poremba, L.~Cincio, A.~T. Sornborger, and P.~J. Coles,
  ``Quantum-assisted quantum compiling,'' \emph{Quantum}, vol.~3, p. 140, 2019.

\bibitem{AmyMMR13}
M.~Amy, D.~Maslov, M.~Mosca, and M.~Roetteler, ``A meet-in-the-middle algorithm
  for fast synthesis of depth-optimal quantum circuits,'' \emph{{IEEE} Trans.
  on {CAD} of Integrated Circuits and Systems}, vol.~32, no.~6, pp. 818--830,
  2013.

\bibitem{MatsumotoA08}
K.~Matsumoto and K.~Amano, ``Representation of quantum circuits with clifford
  and $\pi/8$ gates,'' \emph{arXiv:0806.3834}, 2008.

\bibitem{WilleSOD13}
R.~Wille, M.~Soeken, C.~Otterstedt, and R.~Drechsler, ``Improving the mapping
  of reversible circuits to quantum circuits using multiple target lines,'' in
  \emph{Proceedings of 18th Asia and South Pacific Design Automation
  Conference}, 2013, pp. 145--150.

\bibitem{MaslovFM08}
D.~Maslov, S.~M. Falconer, and M.~Mosca, ``Quantum circuit placement,''
  \emph{{IEEE} Trans. on {CAD} of Integrated Circuits and Systems}, vol.~27,
  no.~4, pp. 752--763, 2008.

\bibitem{SaeediWD11}
M.~Saeedi, R.~Wille, and R.~Drechsler, ``Synthesis of quantum circuits for
  linear nearest neighbor architectures,'' \emph{Quantum Information
  Processing}, vol.~10, no.~3, pp. 355--377, 2011.

\bibitem{Siraichi+18}
M.~Y. Siraichi, V.~F.~d. Santos, S.~Collange, and F.~M.~Q. Pereira, ``Qubit
  allocation,'' in \emph{Proceedings of the International Symposium on Code
  Generation and Optimization}.\hskip 1em plus 0.5em minus 0.4em\relax ACM,
  2018, pp. 113--125.

\bibitem{Venturelli+18}
D.~Venturelli, M.~Do, E.~Rieffel, and J.~Frank, ``Compiling quantum circuits to
  realistic hardware architectures using temporal planners,'' \emph{Quantum
  Science and Technology}, vol.~3, no.~2, p. 025004, 2018.

\bibitem{Murali+19}
P.~Murali, J.~M. Baker, A.~Javadi{-}Abhari, F.~T. Chong, and M.~Martonosi,
  ``Noise-adaptive compiler mappings for noisy intermediate-scale quantum
  computers,'' in \emph{Proceedings of the 24th International Conference on
  Architectural Support for Programming Languages and Operating Systems}.\hskip
  1em plus 0.5em minus 0.4em\relax ACM, 2019, pp. 1015--1029.

\bibitem{ZulehnerPW18}
A.~Zulehner, A.~Paler, and R.~Wille, ``Efficient mapping of quantum circuits to
  the {IBM} {QX} architectures,'' in \emph{Proceedings of the Design,
  Automation {\&} Test in Europe Conference {\&} Exhibition}, 2018, pp.
  1135--1138.

\bibitem{LiDX19}
G.~Li, Y.~Ding, and Y.~Xie, ``Tackling the qubit mapping problem for nisq-era
  quantum devices,'' in \emph{Proceedings of the 24th International Conference
  on Architectural Support for Programming Languages and Operating
  Systems}.\hskip 1em plus 0.5em minus 0.4em\relax ACM, 2019, pp. 1001--1014.

\bibitem{ChildsSU19}
A.~M. Childs, E.~Schoute, and C.~M. Unsal, ``Circuit transformations for
  quantum architectures,'' in \emph{Proceedings of the 14th Conference on the
  Theory of Quantum Computation, Communication and Cryptography}, 2019, pp.
  3:1--3:24.

\bibitem{CowtanDDKSS19}
A.~Cowtan, S.~Dilkes, R.~Duncan, A.~Krajenbrink, W.~Simmons, and S.~Sivarajah,
  ``On the qubit routing problem,'' in \emph{Proceedings of the 14th Conference
  on the Theory of Quantum Computation, Communication and Cryptography}, 2019,
  pp. 5:1--5:32.

\bibitem{Zhou+19}
X.~{Zhou}, S.~{Li}, and Y.~{Feng}, ``Quantum circuit transformation based on
  simulated annealing and heuristic search,'' \emph{IEEE Trans. on CAD of
  Integrated Circuits and Systems}, 2020.

\bibitem{Almeida+19}
A.~A. de~Almeida, G.~W. Dueck, and A.~C. da~Silva, ``Finding optimal qubit
  permutations for {IBM}'s quantum computer architectures,'' in
  \emph{Proceedings of the 32nd Symposium on Integrated Circuits and Systems
  Design}.\hskip 1em plus 0.5em minus 0.4em\relax ACM, 2019, p.~13.

\bibitem{Barenco+95}
A.~Barenco, C.~H. Bennett, R.~Cleve, D.~P. DiVincenzo, N.~Margolus, P.~Shor,
  T.~Sleator, J.~A. Smolin, and H.~Weinfurter, ``Elementary gates for quantum
  computation,'' \emph{Phys. Rev. A}, vol.~52, pp. 3457--3467, Nov 1995.

\bibitem{Rodney14}
R.~Van~Meter, \emph{Quantum Networking}.\hskip 1em plus 0.5em minus 0.4em\relax
  John Wiley \& Sons, 2014.

\bibitem{CordellaFSV04}
L.~P. Cordella, P.~Foggia, C.~Sansone, and M.~Vento, ``A (sub)graph isomorphism
  algorithm for matching large graphs,'' \emph{{IEEE} Trans. Pattern Anal.
  Mach. Intell.}, vol.~26, no.~10, pp. 1367--1372, 2004.

\bibitem{qiskit}
G.~Aleksandrowicz, T.~Alexander, P.~Barkoutsos, L.~Bello, Y.~Ben-Haim,
  D.~Bucher, F.~Cabrera-Hern{\'a}ndez, J.~Carballo-Franquis, A.~Chen, C.~Chen
  \emph{et~al.}, ``Qiskit: An open-source framework for quantum computing,''
  \emph{Accessed on: March 16, 2019}, 2019.

\bibitem{Cowtan_2020}
A.~Cowtan, S.~Dilkes, R.~Duncan, W.~Simmons, and S.~Sivarajah, ``Phase gadget
  synthesis for shallow circuits,'' \emph{Electronic Proceedings in Theoretical
  Computer Science}, vol. 318, p. 213–228, May 2020.

\bibitem{Siraichi+19}
M.~Y. Siraichi, V.~F.~d. Santos, C.~Collange, and F.~M. Q.~a. Pereira, ``Qubit
  allocation as a combination of subgraph isomorphism and token swapping,''
  \emph{Proc. ACM Program. Lang.}, vol.~3, no. OOPSLA, Oct. 2019.

\bibitem{Lin+19}
J.~X. Lin, E.~R. Anschuetz, and A.~W. Harrow, ``Using spectral graph theory to
  map qubits onto connectivity-limited devices,'' \emph{arXiv:1910.11489},
  2019.

\end{thebibliography}


%
\begin{IEEEbiography}[{\includegraphics[width=1in,height=1.25in,clip,keepaspectratio]{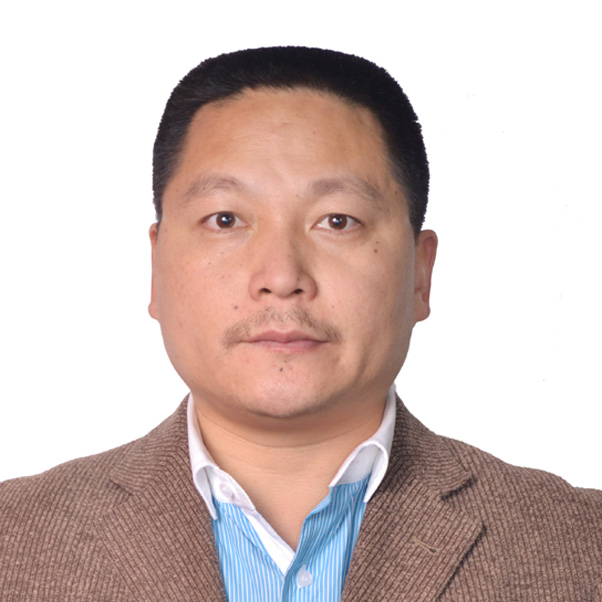}}]{Sanjiang Li}
 received his B.Sc. and Ph.D. degrees in mathematics from Shaanxi Normal University, in 1996, and Sichuan University, in 2001, respectively. He is a professor in Centre for Quantum Software and Information,  University of Technology Sydney (UTS). Before joining UTS, he worked in the Department of Computer Science and Technology, Tsinghua University from 2001 to 2008. His research interests are mainly in knowledge representation and artificial intelligence. 
\end{IEEEbiography}
\begin{IEEEbiography}[{\includegraphics[width=1in,height=1.25in,clip,keepaspectratio]{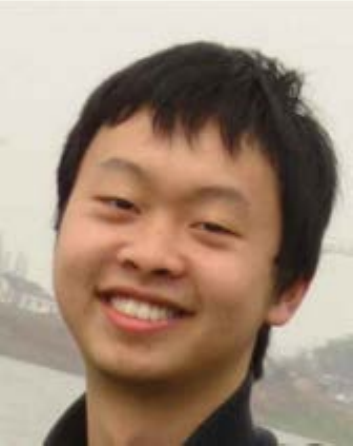}}]{Xiangzhen Zhou}
received the BE degree (2010) in electronic engineering from Nanjing Normal University, China. Currently, he is a PhD candidate at Southeast University, China and also a visiting student in Centre for Quantum Software and Information, University of Technology Sydney. His research interests include quantum computing and quantum circuit optimisation.
\end{IEEEbiography}

\begin{IEEEbiography}[{\includegraphics[width=1in,height=1.25in,clip,keepaspectratio]{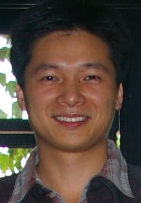}}]{Yuan Feng}
received the BS and PhD degrees from the Department of Applied Mathematics and the Department of Computer Science and Technology, Tsinghua University, in 1999 and 2004, respectively. He is currently a professor at the Centre for Quantum Software and Information (QSI), University of Technology Sydney (UTS), Australia. His research interests include quantum programming theory, quantum information and quantum computation, and probabilistic systems.
\end{IEEEbiography}




  \section*{Acknowledgements}
This work was partially supported by the National Key R\&D Program of China (Grant No. 2018YFA0306704) and the Australian Research Council (Grant No. DP180100691).


\end{document}